\documentclass[12pt]{article}
\pdfoutput=1  
\usepackage[margin=0.95 in]{geometry}
\usepackage{amsmath}
\usepackage{amssymb,amsfonts}
\usepackage[all]{xy}
\usepackage{graphicx}
\usepackage[utf8x]{inputenc}
\usepackage{amsmath}
\usepackage{amssymb}
\usepackage{float}
\usepackage{array}
\usepackage{tikz}
\usepackage{mathtools}
\usepackage{mathrsfs} 
\usepackage[hidelinks]{hyperref}
\usepackage{cite}
\usepackage{tikz-cd}
\usepackage{graphicx}
\usepackage{bm}
\usepackage{rotating}
\usepackage{array}
\usepackage{xcolor}
\usepackage{amsmath,amssymb}
\usepackage{mathrsfs}
\usepackage{graphicx}
\usepackage{color}
\usepackage{subfigure}
\usepackage{fancyhdr}
\usepackage{multirow}
\usepackage{float}
\usepackage{epsfig}
\usepackage{amsfonts}
\usepackage{bm}

\newcommand{\ba}{\begin{eqnarray}}
\newcommand{\ea}{\end{eqnarray}}

\numberwithin{equation}{section}
\numberwithin{equation}{section}
\numberwithin{table}{section}

\begin{document}

\vspace*{3cm}{}
	
	\noindent
	{\LARGE \bf  Traversable wormholes in $f(R)$-massive gravity}
	\vskip .4cm
	\noindent
	\linethickness{.04cm}
	\line(10,0){467}
	\vskip 1cm
	\noindent
	{\large \bf  Takol Tangphati$^{1,*}$, Auttakit Chatrabhuti$^{1,**}$, Daris Samart$^{2,\dagger}$, \\Phongpichit Channuie$^{3,4,5,6,\ddagger}$}
\vskip 0.5cm
{\em 
\noindent
$^1$Department of Physics, Faculty of Science, Chulalongkorn University, \\Bangkok 10330, Thailand\\
$^2$Department of Physics, Faculty of Science, Khon Kaen University, \\Khon Kaen, 40002, Thailand\\
$^3$School of Science, Walailak University, Nakhon Si Thammarat, 80160, Thailand \\
$^4$College of Graduate Studies, Walailak University, Nakhon Si Thammarat, 80160, Thailand\\
$^5$Research Group in Applied, Computational and Theoretical Science (ACTS), \\Walailak University, Nakhon Si Thammarat, 80160, Thailand\\
$^6$Thailand Center of Excellence in Physics, Ministry of Higher Education, Science, \\Research and Innovation, Bangkok 10400, Thailand\\
{\rm Email:}\,$^*${\rm takoltang@gmail.com},\,$^{**}${\rm auttakit@sc.chula.ac.th},\,$^{\dagger}${\rm darisa@kku.ac.th},\,$^{\ddagger}${\rm channuie@gmail.com}}

\vskip 1cm

\noindent {\sc Abstract: }  In this work, the study of traversable wormholes in $f(R)$-massive gravity with the function $f(R)=R+\alpha_{1} R^{n}$, where $\alpha_{1}$ and $n$ are arbitrary constants, is considered. We choose the shape function of the form $b(r)=r \exp(-\alpha(r−r_{0}))$ with $\alpha$ and $r_{0}$ being an arbitrary constant and a radius of the wormhole throat, respectively. Here $\alpha$ affects the radius of curvature of the wormhole. We consider a spherically symmetric and static wormhole metric and derive field equations. Moreover, we visualize the wormhole geometry using embedding diagrams. Furthermore, we check the null, weak, dominant and strong energy conditions at the wormhole throat with a radius $r_{0}$ invoking three types of redshift functions, $\Phi={\rm constant},\,\gamma_{1}/r,\,\log(1+\gamma_{2}/r)$ with $\gamma_{1}$ and $\gamma_{2}$ are arbitrary real constants. We also compute the volume integral quantifier to calculate the amount of the exotic matter near the constructed wormhole throat.
	
\vskip 1cm

	
\newpage
\setcounter{tocdepth}{2}
\tableofcontents

\section{Introduction}
Two-relevant passages written by C. Sagan \cite{CSagan1985} drive us to search for a new perspective on interstellar traveling. The passages dissent an interstellar traveling through black holes and Schwarzschild wormholes. The later is due to the collapse too quickly for anything to cross from one end to the other and it would be possible only if some exotic matters with negative energy density could be used to stabilize them. Wormholes that could be crossed in both directions are known as traversable wormholes \cite{Morris:1988cz}. In the literature, many authors have intensively studied various aspects of traversable wormhole (TW) geometries, with and without invoking exotic matters. Very recently, Casimir energy is successfully used as a potential source to generate various types of traversable wormholes, see e.g., \cite{Garattini:2019ivd,Jusufi:2020rpw}. Introducing exotic matters, one expects that traversable wormhole solutions violate the energy conditions at the wormhole throat.

It was found that modified theories of gravity such as $f(R)$ theory also play an important role in studying wormholes. In Ref.\cite{Lobo2009}, the authors investigated
traversable wormholes using the framework of $f(R)$ gravity. Here the factors responsible
for the violation of null energy condition and supporting the existence of wormholes have been analyzed. In addition, wormhole solutions for different shape functions have been examined. Author of Ref.\cite{Bronnikov:2010tt} discussed the existence of wormholes in scalar tensor theory and $f(R)$ gravity. Null
and weak energy conditions for wormholes with constant shape and redshift functions in $f(R)$ gravity can be found in Ref.\cite{Saeidi:2011zz}. Dynamical wormholes in $f(R)$ gravity was developed in Ref.\cite{Bahamonde:2016ixz}, while wormhole solutions using different types of shape functions in $f(R)$ gravity were carried out in Ref.\cite{Kuhfittig:2018vdg}. The study of static and spherically symmetric traversable wormholes in the presence of cosmological constant was explored in Ref.\cite{Lemos2003}. There were various types of wormhole solutions in modified theories of gravity, e.g. Einstein-Gauss-Bonnet gravity \cite{Maeda:2008nz},  $f(R, \phi)$ gravity \cite{Zubair:2017oir}, both $f(R)$ and $f(R, T)$ theories \cite{Sahoo2018,Samanta:2018hbw}, Born-Infeld gravity \cite{Shaikh2018}, Eddington-inspired Born-Infeld gravity \cite{Harko2015}, and even in non-commutative geometry \cite{Jamil2014,Rahaman2012a}. Moreover, there exist many other models of wormholes different from those mentioned above, see e.g. \cite{Boehmer2012,Kord2015,Clement1984,Mehdizadeh2017,Jusufi2016,
Jusufi2018a,Jusufi2019,Dai2019,Teo1998,Ovgun2019a,Shaikh2016,
Rahaman2016,Montelongo2011,Hochberg1998,Visser2003}. These include the constructions of traversable wormholes in various types of $f(R)$ gravity \cite{Golchin:2019qch}.

Among various modified theories of gravity, the de Rham-Gabadadze-Tolley (dRGT) massive theory \cite{deRham:2010kj,deRham:2010ik,deRham:2014zqa} is targeted as one of the compelling scenarios when studying the universe in the cosmic scale. The  applications of dRGT massive gravity to study the exotic objects, e.g., black holes, already appeared in the literature \cite{Ghosh:2015cva,Chabab:2019mlu, Boonserm:2017qcq}. In the present work, we construct traversable wormholes by engaging the original massive gravity with $f(R)$ gravity. Here we consider $f(R)=R+\alpha_{1} R^{n}$ with $\alpha_{1}$ and $n$ being arbitrary constants. The structure of the present work is as follows: In Sec.\ref{sec2}, we consider the action of $f(R)$-dRGT massive gravity and derive field equations of the underlying theory. In Sec.\ref{sec3}, we choose the wormhole shape function $b(r)=r \exp(-\alpha(r−r_{0}))$ with $\alpha$ and $r_{0}$ being an arbitrary constant and a radius of the wormhole throat, respectively, and consider three types of the redshift function, $\Phi={\rm constant},\,\gamma_{1}/r,\,\log(1+\gamma_{2}/r)$ with $\gamma_{1}$ and $\gamma_{2}$ are arbitrary real constants. Moreover, we visualize the wormhole geometry using embedding diagrams. In Sec.\ref{sec4}, we check the null, weak, dominant and strong conditions at the wormhole throat. In addition, we compute the total  amount of averaged  null energy condition (ANEC) violating matter in the space-time in Sec.\ref{sec5}. Finally, we conclude our findings in the last section. Note that for numerical evaluations we use the geometrical units such that $G=1=c$. 

\section{Field equations}
\label{sec2}
In this section, in the context of $f(R)$ theories, we consider the dRGT theory of massive gravity. The action of the dRGT model on the manifold $\mathcal{M}$ with the presence of $f(R)$ takes the form
\begin{equation}
    S = \int d^4 x \sqrt{-g} \bigg(\frac{1}{16\pi G}\Big[ f(R) + m^2_g \mathcal{U}(g,\phi^a) \Big]\bigg) + \int d^4 x \sqrt{-g} L_{m},
    \label{action_charge_dRGT}
\end{equation}
where $g$ is a determinant of the metric tensor $g_{\mu\nu}$, $f(R)$ is an arbitrary function of $R$, $\mathcal{U}$ is a potential for the graviton which modifies the gravitational sector with the parameter $m_{g}$ interpreted as graviton mass, and $L_{m}$ is the matter field sector. The potential of the massive gravity, $\mathcal{U}$ is defined by 
\begin{eqnarray}
    \mathcal{U} = \mathcal{U}_2 + \alpha_3 \mathcal{U}_3 + \alpha_4 \mathcal{U}_4\,.\label{alpha34}
\end{eqnarray} 
The functions $\mathcal{U}_{2},\,\mathcal{U}_{3}$ and $\mathcal{U}_{4}$ are given by
\begin{eqnarray}
    \mathcal{U}_2 &=& [\mathcal{K}]^2 - [\mathcal{K}^2], \nonumber \\
    \mathcal{U}_3 &=& [\mathcal{K}]^3 - 3[\mathcal{K}][\mathcal{K}^2] + 2 [\mathcal{K}^3], \nonumber \\
    \mathcal{U}_4 &=& [\mathcal{K}]^4 - 6[\mathcal{K}]^2[\mathcal{K}^2] + 8[\mathcal{K}][\mathcal{K}^3] + 3[\mathcal{K}^2]^2 -     6[\mathcal{K}^4], \nonumber \\
    \mathcal{{K}^{\mu}}_{\nu} &=& \delta^{\mu}_{\nu} - \sqrt{g^{\mu\lambda} \mathcal{F}_{ab} \partial_{\lambda}\phi^a \partial_{\nu}\phi^b },
    \label{kappa_dRGT}
\end{eqnarray}
where a bracket $[\quad]$ represents the trace of the rank-two tensor, $\mathcal{K}_\nu^\mu$. Moreover, the fiducial metric, $\mathcal{F}_{ab}$ is chosen such that 
\begin{equation}
    \mathcal{F}_{ab} = 
        \begin{pmatrix}
        0 & 0 & 0 & 0 \\
        0 & 0 & 0 & 0 \\
        0 & 0 & k^2 & 0 \\
        0 & 0 & 0 & k^2 \text{sin}^2 \theta \\
        \end{pmatrix},
    \label{fd_matrix}
\end{equation}
where $k$ is a positive constant and the unitary gauge is used as 
\begin{equation}
	\phi^a = x^{\mu} \delta^a_{\mu}\,.
\end{equation}
In addition, the parameters $\alpha_{3,4}$ are the parameters of the dRGT theory and we will relate these parameters with the graviton mass in the latter. To obtain the Einstein field equation of the dRGT massive gravity with the Maxwell field, we vary the gravitational action in Eq.(\ref{action_charge_dRGT}) with respect to the metric, $g^{\mu\nu}$ to yield
\begin{equation}
f'(R)R_{\mu\nu}-\frac{1}{2}f(R) g_{\mu\nu}-\nabla_{\mu}\nabla_{\nu}F+g_{\mu\nu}\Box F= - m_g^2 X_{\mu \nu} +  8\pi G\Big(T^{(m)}_{\mu\nu}\Big),
\label{Einstein_dRGT_charge}
\end{equation}
where $T^{(m)}_{\mu\nu}$ is the energy-momentum tensor of the matter field, $F = F(R)\equiv df(R)/dR$ and $\Box F=g^{\mu\nu}\nabla_{\mu}\nabla_{\nu} F$. Here $X_{\mu \nu}$ is the dRGT massive gravity tensor. One may write
\begin{eqnarray}
\frac{m_g^2}{8\pi G}\,X_{\mu\nu} &=& -\Big( \rho^{(g)} + p_t^{(g)}\Big)u_\mu u_\nu - p_t^{(g)} g_{\mu\nu} - \Big( p_r^{(g)} - p_t^{(g)}\Big)\chi_\mu \chi_\nu\,. 
\label{EMT-dRGT}
\end{eqnarray}
The energy momentum tensor for the matter source of the wormholes is $T^{(m)}_{\mu\nu} = -\frac{2}{\sqrt{-g}}\frac{\partial \sqrt{-g} L_{m}}{\partial g^{\mu\nu}}$, which is
defined in terms of the principle pressures as
\begin{eqnarray}
T^{(m)}_{\mu\nu} = \Big( \rho + P_t\Big)u_\mu u_\nu + P_t\, g_{\mu\nu}+ \Big( P_r - P_t\Big)\chi_\mu \chi_\nu\,, 
\label{EMT-matter}
\end{eqnarray}
where the $u_\mu$ is a time-like unit vector and the $\chi_\mu$ is the spacelike unit vector orthogonal to the $u_\mu$ with the normalization condition $u_\mu u^\mu = -1$ and $\chi_\mu \chi^\mu =1$. The field equation (\ref{Einstein_dRGT_charge}) can be rewritten in the following form
\begin{eqnarray}
G_{\mu\nu}=R_{\mu\nu}-\frac{1}{2}Rg_{\mu\nu}= - m_g^2 X_{\mu \nu} +  8\pi G\Big(T_{\mu\nu}^{(f(R))}+T^{(m)}_{\mu\nu}\Big),
\label{fRdRGT_charge}
\end{eqnarray}
where
\begin{eqnarray}
8\pi G T_{\mu\nu}^{(f(R))}=\frac{1}{2}g_{\mu\nu}(f(R)-R)+\nabla_{\mu}\nabla_{\nu}F-g_{\mu\nu}\Box F +(1-F)R_{\mu\nu}.
\label{fRdRGT_charge}
\end{eqnarray}
The $X_{\mu\nu}$ tensor from the dRGT massive gravity is defined by
\begin{eqnarray}
    X_{\mu\nu} &=& \mathcal{K}_{\mu\nu} -\alpha \left[ \big(\mathcal{K}^2 \big)_{\mu\nu} - [\mathcal{K}]\mathcal{K}_{\mu\nu} + \frac12 g_{\mu\nu}\big( [\mathcal{K}]^2 - [\mathcal{K}^2]\big)\right]
    \label{X-munu}\\
    && +\, 3\beta\left[ \big(\mathcal{K}^3\big)_{\mu\nu} - [\mathcal{K}]\big( \mathcal{K}^2\big)_{\mu\nu} + \frac12 \mathcal{K}_{\mu\nu}\big( [\mathcal{K}]^2 - [\mathcal{K}^2]\big) -\frac16 g_{\mu\nu}\big( [\mathcal{K}]^3 - 3[\mathcal{K}][\mathcal{K}^2] + 2[\mathcal{K}^3]\big)\right] ,
    \nonumber
\end{eqnarray}
where the parameters $\alpha$ and $\beta$ are related to $\alpha_{3,4}$ from the action in Eq.(\ref{alpha34}) via
\begin{eqnarray}
    \alpha = 1 + 3\alpha_3\,,\qquad \beta = \alpha_3 + 4\alpha_4.
\end{eqnarray}
According to the definition of the energy momentum tensor of the dRGT massive gravity in Eq.(\ref{EMT-dRGT}), one can compute the $\rho^{(g)}$ and $p_{r,\perp}^{(g)}$ from the $m_g^2X_{\mu\nu}$ term in Eq.(\ref{X-munu}) directly. They are given by \cite{Burikham:2016cwz,Kareeso:2018xum,Panpanich:2018cxo}
\begin{eqnarray}
    \rho^{(g)}(r) &\equiv& \frac{m_g^2}{8\pi G}{X^t}_t 
    = -\frac{1}{8\pi G}\left( \frac{2 \gamma -  \Lambda r}{r} \right),
    \label{rho-g}\\
    p_r^{(g)}(r) &\equiv& -\frac{m_g^2}{8\pi G}{X^r}_r = \frac{1}{8\pi G}\left( \frac{2 \gamma -  \Lambda r}{r} \right),
    \label{p-g-r}\\
    p_\perp^{(g)}(r) &\equiv& -\frac{m_g^2}{8\pi G}X_{\theta,\phi}^{\theta,\phi} 
    = \frac{1}{8\pi G}\left( \frac{ \gamma -  \Lambda r}{r} \right),
    \label{p-g-theta}
\end{eqnarray}
where $\Lambda$ is the effective cosmological constant, and $\gamma$ is a new parameter, and they are linear combinations of the parameters in the dRGT massive gravity via the following relations:
\begin{eqnarray}
    \Lambda \equiv -3m_g^2(1+ \alpha + \beta), \quad \gamma \equiv -m_g^2k(1 + 2\alpha + 3\beta). 
    \label{para-fr}
\end{eqnarray}
In addition, it has been shown that the energy-momentum tensor of the massive gravity, $T_{\mu\nu}^{(g)}$ exhibits its behavior like anisotropic dark energy i.e., $p_r^{(g)} = -\rho^{(g)}$ see Refs. \cite{Burikham:2016cwz,Kareeso:2018xum,Panpanich:2018cxo} for detail discussions and applications. The total energy momentum tensor of the Einstein field equation in the mixed tensor form can be expressed in the diagonal matrices as $T_{\mu\nu}={\rm diag}\,\big(-\rho^{(g)}-\rho,\,p_r^{(g)}+P_{r},\,p_t^{(g)}+P_{t},\,p_t^{(g)}+ P_{t}\big)$. Moreover, the conservation of the total energy momentum tensor is hold as well as for the perfect fluid and the massive graviton parts as $\nabla^\mu T_{\mu\nu}^{(g)} = 0,\,\nabla^\mu T_{\mu\nu}^{(m)} = 0$.

\section{Wormhole geometry}
\label{sec3}
In order to construct the traversable wormholes, we consider a static and spherically symmetric Morris-Thorne traversable wormhole given by \cite{Morris:1988cz}
\begin{eqnarray}
    ds^2 = - e^{2 \Phi(r)}dt^2 + \frac{dr^2}{1 - \frac{b(r)}{r}} + r^2 d\Omega^2,
    \label{metric_traversable}
\end{eqnarray}
where $d\Omega^2\equiv d\theta^2 + \text{sin}^2 \theta d \phi^2$. The characteristic of the wormhole is determined by the wormhole shape function $b(r)$ and the red shift function $\Phi(r)$. Since we consider a spherically symmetric metric, one may consider an equatorial slice $\theta=\pi/2$ and for a fixed moment of time i.e. $t={\rm constant}$, the metric (\ref{metric_traversable}) simply reduces to
\begin{eqnarray}
    ds^2 = \frac{dr^2}{1 - \frac{b(r)}{r}} + r^2 d\phi^2.
    \label{metric_traversable1}
\end{eqnarray}
In terms of cylindrical coordinates, The equation (\ref{metric_traversable1}) is equivalent to
\begin{eqnarray}
    ds^2 = dz^2 + dr^2+ r^2 d\phi^2,
    \label{metricEu}
\end{eqnarray}
Because the embedded surface in the three-dimensional Euclidean space is characterized by $z=z(r)$, the metric (\ref{metricEu}) of the surface can be recast to yield
\begin{eqnarray}
    ds^2 = \left(1+\bigg(\frac{dz}{dr}\bigg)^{2}\right)dr^2+ r^2 d\phi^2,
    \label{metricEu1}
\end{eqnarray}
Comparing Eq.(\ref{metric_traversable1}) and Eq.(\ref{metricEu1}), we obtain
\begin{eqnarray}
    \frac{dz}{dr} = \pm \Big(\frac{r}{b(r)}-1\Big)^{-1/2}.
    \label{traversable1}
\end{eqnarray}
\begin{figure}[!h]
    \centering
    \includegraphics[width =8.3cm]{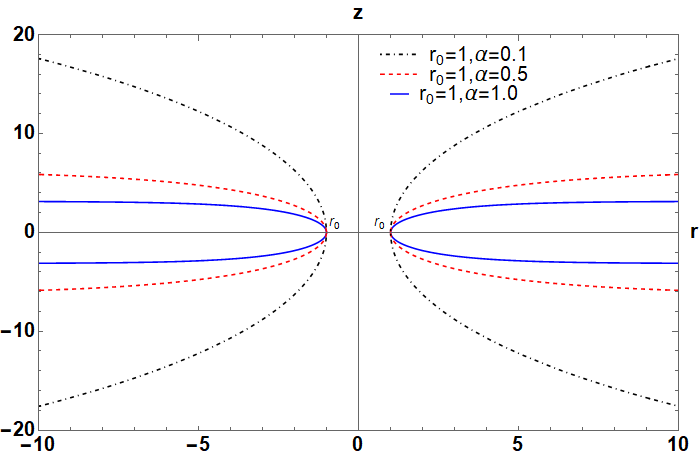}
     \includegraphics[width =8.3cm]{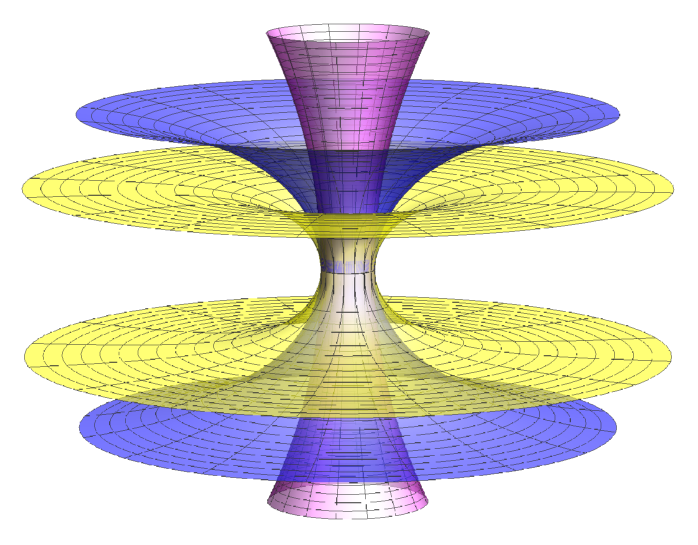}
    \caption{Plots show embedding diagrams of the metric (\ref{metric_traversable1}) for slices $t= {\rm const,}\,\theta=\pi/2$. Left-panel shows the 2-dimensional diagram of the constructed traversable wormhole using $r_0 = 1.0, \text{ and } \alpha = 0.1$ (a black dot-dashed line), $r_0 = 1.0, \text{ and } \alpha = 0.5$ (a red dot line) and $r_0 = 1.0, \text{ and } \alpha = 1.0$ and (a blue solid line). Right-panel displays 3-dimensional diagram of the constructed traversable wormholes using the same sets of parameters.}
    \label{2D1}
\end{figure}

\begin{figure}[!h]
    \centering
    \includegraphics[width =8.3cm]{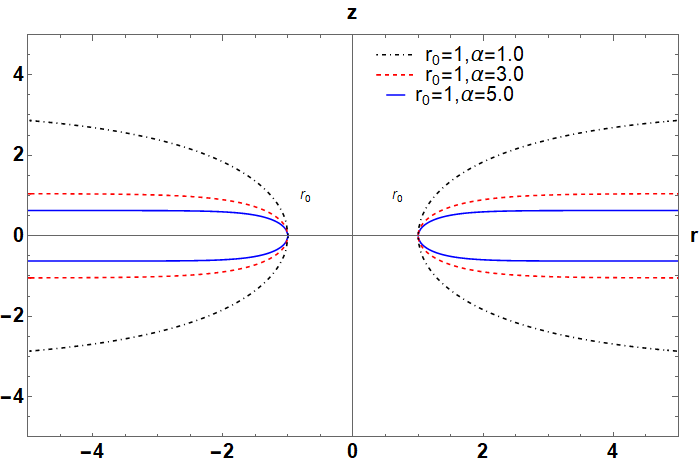}
     \includegraphics[width =8.3cm]{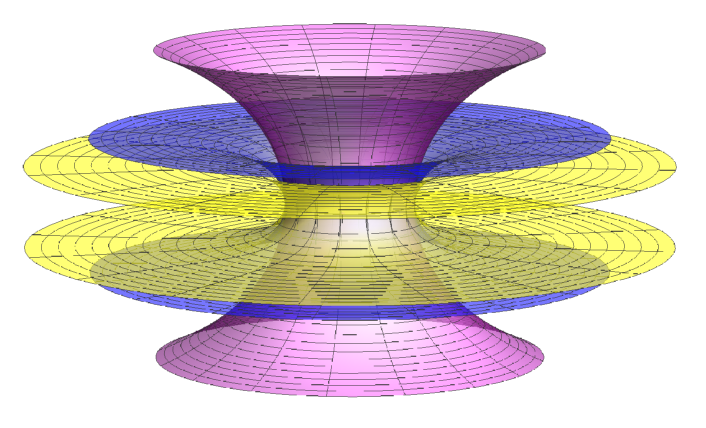}
    \caption{Plots show embedding diagrams of the metric (\ref{metric_traversable1}) for slices $t= {\rm const,}\,\theta=\pi/2$. Left-panel shows the 2-dimensional diagram of the constructed traversable wormhole using $r_0 = 1.0, \text{ and } \alpha = 1.0$ (a black dot-dashed line), $r_0 = 1.0, \text{ and } \alpha = 3.0$ (a red dot line) and $r_0 = 1.0, \text{ and } \alpha = 5.0$ and (a blue solid line). Right-panel displays 3-dimensional diagram of the constructed traversable wormholes using the same three sets of parameters.}
    \label{2D2}
\end{figure}

At this point, the geometry of the wormhole solutions must follow the condition at the wormhole throat, i.e. $b(r=r_{0})=r_{0}$, where $r_{0}$
denotes the radius of the wormhole throat. Notice that the embedded surface is vertical at the throat, i.e. $dz/dr\rightarrow \infty$ at $r=r_{0}$, and the wormhole space is asymptotically flat as $r\rightarrow \infty$, i.e. $dz/dr\rightarrow 0$ at $r\rightarrow \infty$. Moreover, the inverse of the embedding function $r(z)$ must satisfy $d^{2}r/dz^{2}>0$ near or at the wormhole throat, $r_{0}$. This is so-called the flare-out condition. More concretely, we differentiate $dr/dz$ of Eq.(\ref{traversable1}) with respect to $z$, and then we find
\begin{eqnarray}
    \frac{d^{2}r}{dz^{2}} = \frac{b(r)-rb'(r)}{2b(r)^{2}}>0\,,
    \label{flare}
\end{eqnarray}
\begin{figure}[!h]
    \centering
    \includegraphics[width =8.3cm]{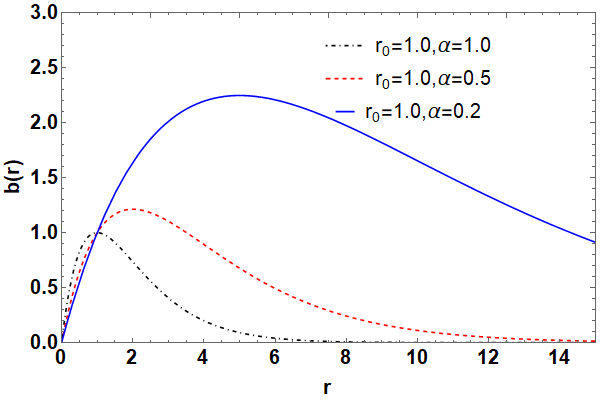}
    \includegraphics[width =8.3cm]{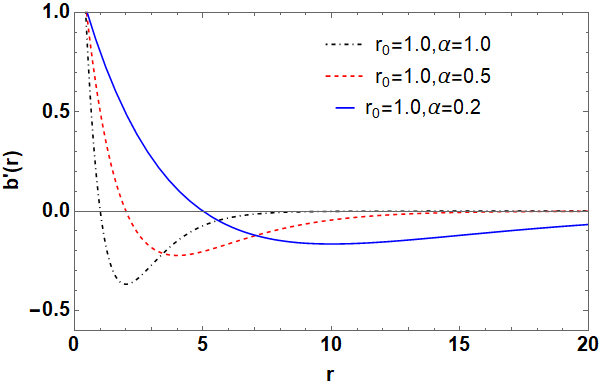}
    \includegraphics[width =8.3cm]{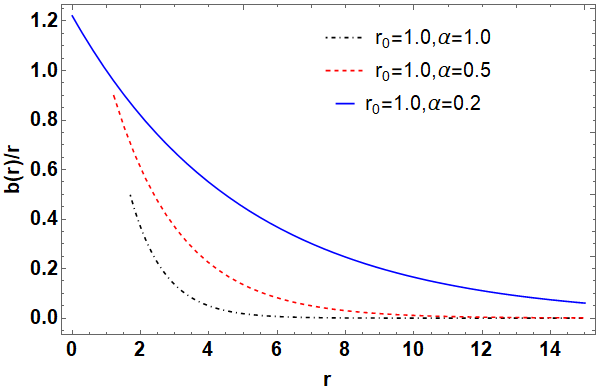}
    \includegraphics[width =8.3cm]{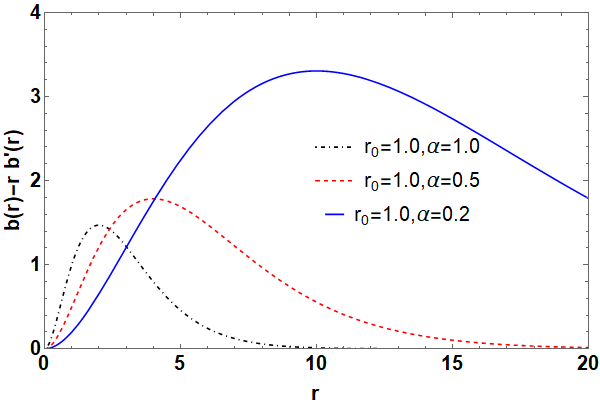}
    \caption{We verify the properties of the shape function introduced in Eq.(\ref{shap1}). The plots show behaviors of the proposed shape function against the requirements given by ($i$)-($v$) using various values of $\alpha=0.1,\,0.5,\,1.0$ and $r_{0}=1$. We find that the shape function of the wormhole is completely satisfied the requirements.}
    \label{br}
\end{figure}

\begin{figure}[!h]
    \centering
    \includegraphics[width =8.3cm]{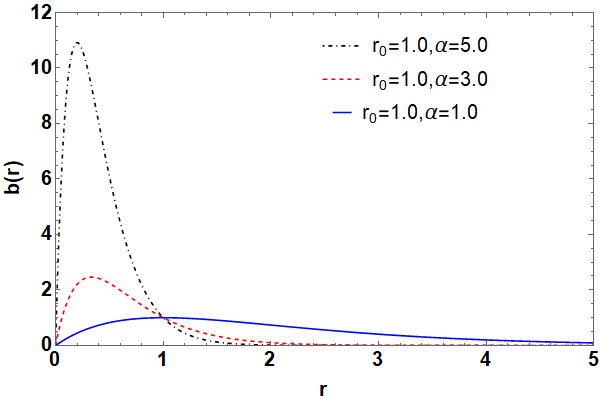}
    \includegraphics[width =8.3cm]{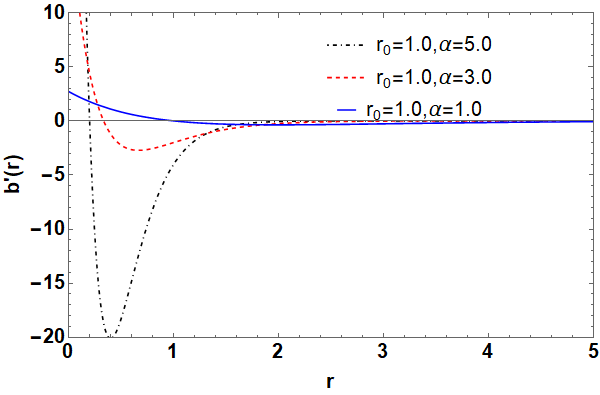}
    \includegraphics[width =8.3cm]{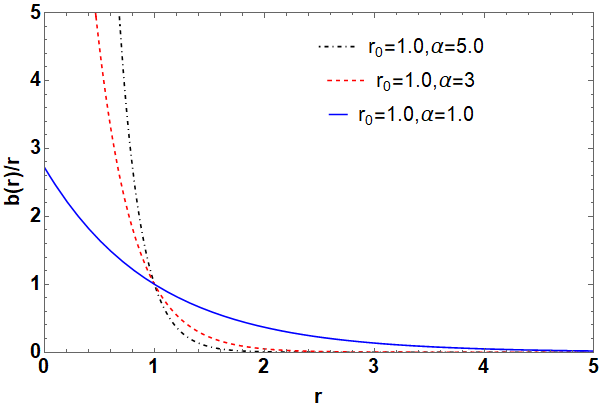}
    \includegraphics[width =8.3cm]{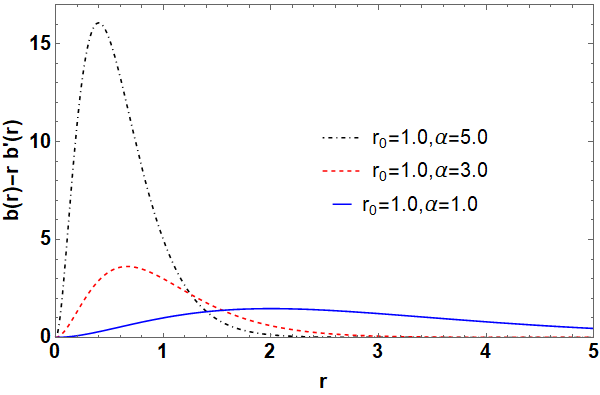}
    \caption{We verify the properties of the shape function introduced in Eq.(\ref{shap1}). The plots show behaviors of the proposed shape function against the requirements given by ($i$)-($v$) using various values of $\alpha=1.0,\,3.0,\,5.0$ and $r_{0}=1$. We find that the shape function of the wormhole is completely satisfied the requirements.}
    \label{br1}
\end{figure}

where a prime denotes derivative with respect to $r$. In order to have a consistent WH construction, the shape function should satisfy the following properties: ($i$) $b(r)/r<1$ for $r>r_{0}$,\,($ii$) $b(r)=r_{0}$ at $r=r_{0}$,\,($iii$) $b(r)/r\rightarrow 0$ as $r\rightarrow \infty$,\,($iv$) $b(r)−b′(r)r >0$ and ($v$) $b'(r)<1$ at $r=r_{0}$. In this work, we choose the shape function $b(r)$ of the form
\begin{eqnarray}
    b(r) = \frac{r}{\exp(\alpha (r - r_0))},\label{shap1}
\end{eqnarray}
where $\alpha$ and $r_{0}$ are an arbitrary constant and a radius of the wormhole throat, respectively. Notice that $\alpha=1$ is considered in Refs.\cite{Samanta:2018hbw,Godani:2019kgy} for constructing traversable wormholes in modified theories of gravity. The properties of the proposed shape function can be displayed in Fig.(\ref{br}). We clearly see that the shape function introduced in Eq.(\ref{shap1}) is nicely satisfied all required properties ($i$)-($v$).

Considering Eq.(\ref{Einstein_dRGT_charge}), the effective field equations for the metric (\ref{metric_traversable}) can be expressed as follows:
\begin{eqnarray}
    \rho &=& \frac{dF(R)}{dr} \left(\frac{r b'(r)+3 b(r)-4 r}{16 \pi  G r^2} - \frac{1}{8 \pi G} \left( 1 - \frac{b(r)}{r} \right) \Phi '(r) \right)+\frac{f(R)}{16 \pi  G}+\frac{2 \gamma  r - \Lambda  r^2}{8 \pi  G r^2} \nonumber \\
    && + F(R) \left(\frac{\left(r \left(-b'(r)\right)-3 b(r)+4 r\right) \Phi '(r)}{16 \pi G r^2} + \frac{1}{8 \pi G} \left( 1 - \frac{b(r)}{r} \right) \left( \Phi''(r) + \Phi'^2(r) \right) \right) \nonumber \\
    && - \frac{1}{8 \pi G} \left( 1 - \frac{b(r)}{r} \right) \frac{d^2 F(R)}{dr^2}, \label{rho1}
   \end{eqnarray}
   \begin{eqnarray}
    P_{r} &=& \frac{d F(R)}{dr} \left(\frac{1}{8 \pi G} \left( 1 - \frac{b(r)}{r} \right) \Phi '(r) + \frac{1}{4 \pi G r} \left( 1 - \frac{b(r)}{r} \right)\right)-\frac{f(R)}{16 \pi  G}+\frac{ \Lambda r^2 - 2 \gamma  r}{8 \pi  G r^2} \nonumber \\
    && + F(R) \bigg( \frac{1}{8 \pi G r^3} \left( r b'(r) - b(r) \right) \left( 1 + \frac{r \Phi'(r)}{2} \right) - \frac{1}{8 \pi G} \left( 1 - \frac{b(r)}{r} \right) \left( \Phi''(r) + \Phi'^2(r) \right), \nonumber \\
    \label{pr1} \\
    P_{t} &=& F(R) \left( \frac{1}{8 \pi G} \left( 1 - \frac{b(r)}{r} \right)\Phi'(r) + \frac{1}{16 \pi G r^3} \left( b(r) + r b'(r) \right) \right) -\frac{f(R)}{16 \pi  G}+\frac{ \Lambda r^2 - 2 \gamma  r}{8 \pi  G r^2} \nonumber \\
    && + \frac{dF(R)}{dr} \left( \frac{1}{8 \pi G} \left( 1 - \frac{b(r)}{r} \right) \Phi'(r) + \frac{1}{16 \pi G r^3} \left( 4r^2 - 3 r b(r) - r^2 b'(r) \right) \right) \nonumber \\
    && + \frac{1}{8 \pi G} \left( 1 - \frac{b(r)}{r} \right) \frac{d^2 F(R)}{dr^2}, \label{pt1}
   \end{eqnarray}
where arguments of $b$ and $\Phi$ are understood, i.e., $b=b(r)$ and $\Phi=\Phi(r)$, a prime denotes $\partial/\partial r$, $\rho$ is the energy density, and $P_{r}$ and $P_{t}$ are the radial and tangential pressures, respectively. It is rather straightforward to calculate $\Box F$ and $R$ to obtain
\begin{eqnarray}
\Box F(R) &=& \left( 1 - \frac{b(r)}{r} \right) \left[ \frac{d^2 F(R)}{dr^2} + \left( - \frac{r b'(r) - b(r)}{2 r^2 (1 - b(r) / r)} + \frac{2}{r} + \Phi'(r) \right)\frac{d F(R)}{dr} \right], \\
R &=& \frac{2b'(r)}{r^2} - \frac{(4r - 3b(r) - r b'(r)) \Phi'(r)}{r^2} - \left(1 - \frac{b(r)}{r} \right) \left( \Phi''(r) + \Phi'^2(r) \right).
\end{eqnarray}
To construct wormholes one may consider specific equations of state for $P_{r}$ or $P_{t}$, or restricted choices for the redshift and shape functions, among others. It is worth noting here that to solve the field equations and energy conditions, the functions $b(r)$, $\Phi(r)$ and $F(R)$ need to be basically fixed. Since there are infinitely many possibilities to fix $b(r),\,Phi(r)$ and $F(R)$, there are countless solutions for $\rho$, $P_{r}$ and $P_{t}$. 

In the next section, we propose traversable wormholes by engaging particular $f(R)=R+\alpha_{1} R^{n}$ gravity with dRGT massivegravity. Without dRGT sector, this type of traversable wormholes $f(R)$ model was explored in Ref.\cite{Godani:2019kgy}. However, one can consider many more possible solutions by considering other choices of the functions $b(r)$, $\Phi(r)$ and $F(R)$.

\section{Energy conditions}\label{sec4}
In order to analyse energy conditions in the constructed wormholes, we consider the null energy condition (NEC), weak energy condition (WEC), strong energy condition (SEC) and dominant energy condition (DEC). We consider first NEC which is defined via $T_{\mu\nu}k^{\mu}k^{\nu} \geq 0$ with $k^{\mu}$ being a null-like vector. Alternately, in terms of the principal pressures, NEC is defined as $\rho+P_{i}\geq 0\,\,\forall i$. We next consider WEC which is defined
as $T_{\mu\nu}U^{\mu}U^{\nu}\geq 0$ with $U^{\mu}$ being a time-like vector. In terms of the principal pressures, it is defined as $\rho>0$; and $\rho+P_{i}\geq 0,\,\,\forall i$. The third one is SEC which is defined as $(T_{\mu\nu}-g_{\mu\nu}T/2)U^{\mu}U^{\nu}\geq 0$ where $T$ is the trace of the stress-energy tensor. In terms of the principal pressures, SEC is defined as $T=-\rho+\sum_{i}P_{i}$ and $\rho+\sum_{i}P_{i}\geq 0,\,\,\forall i$. The last one is DEC which is defined as $T_{\mu\nu}U^{\mu}U^{\nu}\geq 0$, and $T_{\mu\nu}U^{\mu}$ is not space-like. In terms of the principal pressures, $\rho \geq 0$; and $P_{i} \in [-\rho, +\rho],\,\,\forall i$. In summary, in terms of
principal pressures, we examine our results by following Ref.\cite{Samanta:2019tjb} in which these conditions are as follows:
\begin{itemize}
  \item NEC: $\rho+P_{r}\geq 0$\,,\, $\rho+P_{t}\geq 0$;
  \item WEC: $\rho\geq 0\,,\,\rho+P_{r}\geq 0$\,,\, $\rho+P_{t}\geq 0$;
  \item SEC: $\rho+P_{r}\geq 0$\,,\, $\rho+P_{t}\geq 0\,,\,\rho+P_{r}+2P_{t}\geq 0$;
  \item DEC: $\rho\geq 0\,,\,\rho-|P_{r}|\geq 0\,,\,\rho-|P_{t}|\geq 0$.
\end{itemize}
In the next subsection, we consider three types of redshift functions, $\Phi={\rm constant},\,\gamma_{1}/r,\,\log(1+\gamma_{2}/r)$ with $\gamma_{1}$ and $\gamma_{2}$ are arbitrary real constants and then analyse energy conditions of those types of redshift functions.
\begin{figure}[!h]
    \centering
    \includegraphics[width = 8 cm]{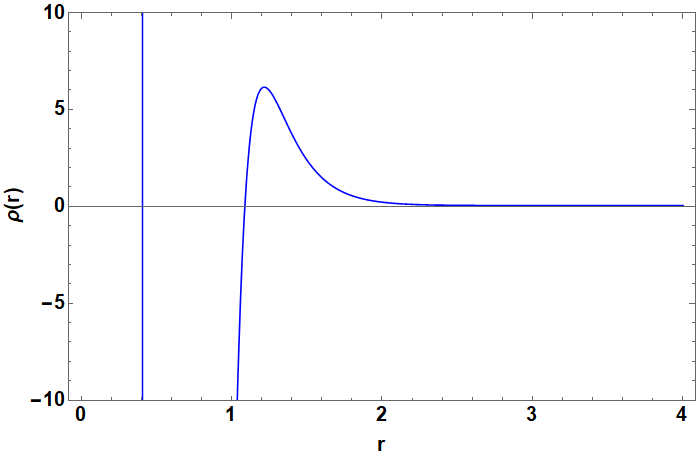}
    \includegraphics[width = 8 cm]{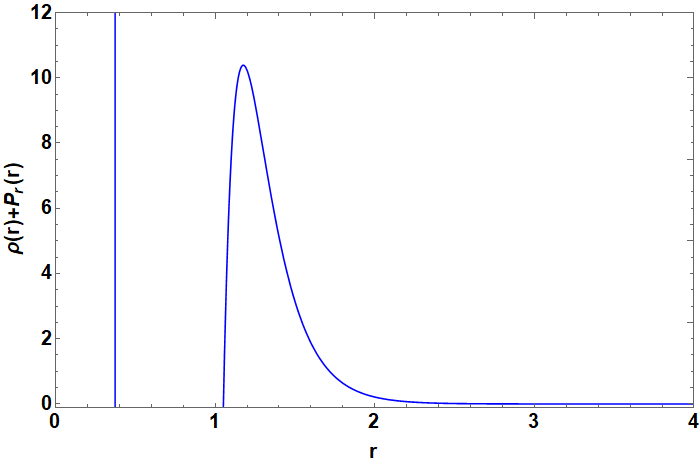}
    \includegraphics[width = 8 cm]{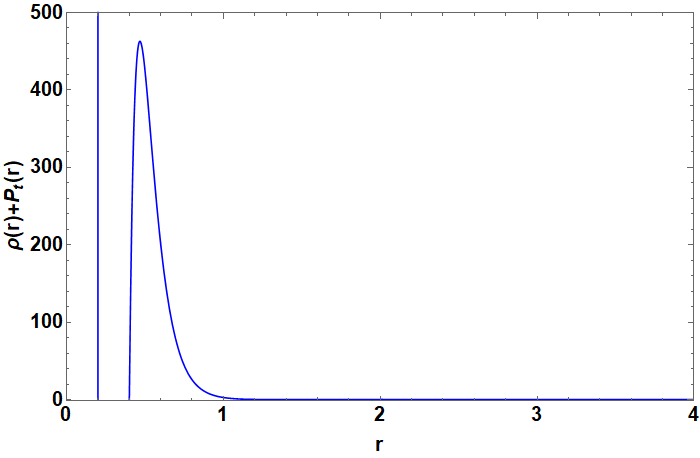}
    \includegraphics[width = 8 cm]{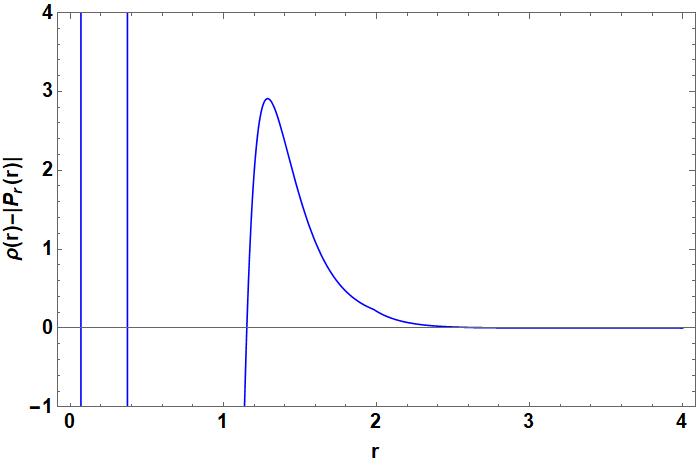}
    \includegraphics[width = 8 cm]{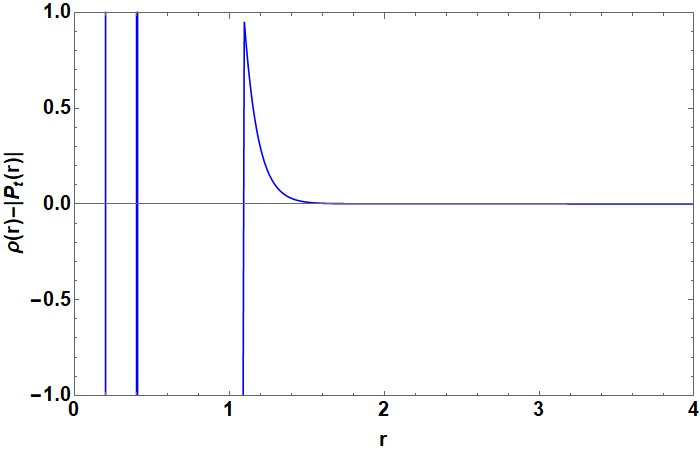}
    \includegraphics[width = 8 cm]{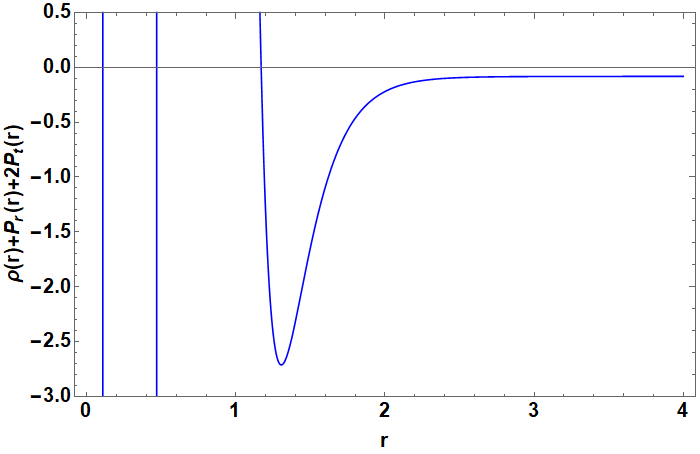}
    \caption{Figures demonstrate the variation of $\rho$, $\rho + P_r$, $\rho + P_t$, $\rho - |P_r|$, $\rho - |P_t|$, $\rho + P_r + 2 P_t$ as a function of $r$ with $\Phi(r) = 1$. We have used $\Lambda = -1.0, \alpha_{1} = 3, n = 2, \alpha = 5.0, r_0 = 1 \text{ and } p = 1.0$.} 
    \label{E1}
\end{figure}
\begin{table}[!h]
	\centering
	\caption{Table shows a summary of energy/pressure conditions for $\Phi(r) = {\rm constant}$,\,$n=2$, $\alpha = 5.0$,$ r_0 = 1, p = 1.0, \gamma = 0.1, \text{ and } \Lambda = -1.0$.}
	\begin{tabular}{|c|c|l|l|}
		\hline\hline
		No.& Terms& $\alpha_{1} < 0$& $\alpha_{1} > 0$ \\
		\hline\hline
		1 & $\rho$ & $\geq 0$, for $r\in(0.40, 1.10] \cup [2.11, \infty)$ & $\geq 0$, for $r \in (0, 0.40] \cup [1.10,\infty)$ \\
		&       & $< 0$, for $r \in (0.00,0.40) \cup (1.10,2.11)$ &  $<0$, for $0.40 < r <1.10$  \\
		\cline{1-4}
		2 & $\rho+P_{r}$ & $\geq0$, for $r\in[0.39,1.05)$&$>0$, for $r\in (0,0.39) \cup (1.05,\infty)$ \\
		&       & $<0$, for $r\in (0,0.39) \cup (1.05,\infty)$ &  $<0$, for $r\in[0.39,1.05)$\\
		\cline{1-4}
		3 & $\rho+P_{t}$ & $\geq0$, for $r\in [0.20,0.41] \cup [1.40, \infty)$& $\geq0$, for $r\in(0.00,0.20) \cup (0.40,\infty)$\\
		&       & $<0$, for $r \in (0,0.20) \cup (0.41,1.40)$ &  $<0$, for $r\in (0.20,0.40)$ \\
		\cline{1-4}
		4 & $\rho+P_{r}+2P_{t}$ & $>0$, for $r \in (0.12, 0.47) \cup (1.18,2.12)$ & $>0$, for $r \in (0,0.11]\cup[0.46, 1.18]$\\
		&       & $<0$, for $r \in (0.47, 1.18)$  & $<0$, for $(0.11,0.46)\cup(1.18,\infty)$ \\
		\cline{1-4}
		5 & $\rho-|P_{r}|$ & $\geq0$, for $r\in[0.48,1.06]$&$>0$, for $r\in[0.83,0.39] \cup [1.17, \infty)$\\
		&       & $<0$, for $r\in (0,0.48) \cup (1.06,\infty)$ &  $<0$, for $r\in (0,0.83) \cup (0.39,1.17)$\\
		\cline{1-4}
		6 & $\rho-|P_{t}|$ & $< 0$, for $\forall r$& $\geq0$, for $r\in(0,0.20] \cup [0.39,0.40]$\\
		&       &  &  \quad $\cup [1.09, \infty)$ \\
		&       &  &  $<0$, for $r \in (0.20,0.39) \cup (0.40,1.09)$ \\
		\hline
	\end{tabular}\label{mod1}
\end{table}

\subsection{$\Phi(r) = \text{constant} = p$}
With the shape function from Eq.~(\ref{shap1}) and the constant red shift function, we find
\begin{eqnarray}
    \rho &=& F'(r) \left(\frac{b'(r)}{16 \pi G r}+\frac{3 b(r)}{16 \pi  G r^2}-\frac{1}{4 \pi  G r}\right) + \left(\frac{b(r)}{8 \pi  G r} - \frac{1}{8 \pi G}\right) F''(r) \nonumber \\ 
    && +\frac{f(r)}{16 \pi G}-\frac{\Lambda }{8 \pi  G}+\frac{\gamma }{4 \pi  G r} \,,\label{rho_1}\\
    P_r &=& F(r) \left(\frac{b'(r)}{8 \pi  G r^2}-\frac{b(r)}{8 \pi  G r^3}\right) + \left(\frac{1}{4 \pi G r}-\frac{b(r)}{4 \pi  G r^2}\right) F'(r) - \frac{f(r)}{16 \pi  G} \nonumber \\
    && + \frac{\Lambda }{8 \pi G} - \frac{\gamma }{4 \pi  G r} \,,\label{P_r_1}\\
    P_t &=& F'(r) \left(-\frac{b'(r)}{16 \pi  G r}-\frac{3 b(r)}{16 \pi  G r^2}+\frac{1}{4 \pi  G r}\right)+F(r) \left(\frac{b'(r)}{16 \pi  G r^2}+\frac{b(r)}{16 \pi  G r^3} \right) \nonumber \\
    && + \left(\frac{1}{8 \pi G} - \frac{b(r)}{8 \pi  G r}\right) F''(r)-\frac{f(r)}{16 \pi  G}+\frac{\Lambda }{8 \pi G}-\frac{\gamma }{8 \pi  G r} \,. \label{P_t_1}
\end{eqnarray}
The combinations of Eqs.~(\ref{rho_1} - \ref{P_t_1}) yield the following relations among $\rho, P_r$, and $P_t$:
\begin{eqnarray}
    \rho + P_r &=& F'(r) \left(\frac{b'(r)}{16 \pi  G r}-\frac{b(r)}{16 \pi  G r^2}\right)+F(r) \left(\frac{b'(r)}{8 \pi G r^2} - \frac{b(r)}{8 \pi  G r^3}\right)\nonumber \\ 
    && +\left(\frac{b(r)}{8 \pi  G r}-\frac{1}{8 \pi  G}\right) F''(r) \,,\label{rho_P_r_1} \\
    \rho + P_t &=& \frac{F(r) b'(r)}{16 \pi  G r^2}+\frac{b(r) F(r)}{16 \pi  G r^3}+\frac{\gamma }{8 \pi  G r} \,, \label{rho_P_t_1}\\
    \rho - |P_r| &=& -\bigg| -\frac{\gamma }{4 G \pi  r}+\frac{\Lambda }{8 G \pi }-\frac{f(r)}{16 G \pi }+F(r)\left(\frac{b'(r)}{8 G \pi  r^2} - \frac{b(r)}{8 G \pi  r^3}\right) \nonumber \\
    && + \left(\frac{1}{4 G \pi r} - \frac{b(r)}{4 G \pi  r^2}\right) F'(r)\bigg| + F'(r) \left(\frac{b'(r)}{16 \pi G r} + \frac{3 b(r)}{16 \pi  G r^2}-\frac{1}{4 \pi  G r}\right) \nonumber \\
    && +\left(\frac{b(r)}{8 \pi G r} - \frac{1}{8 \pi  G}\right) F''(r) +\frac{f(r)}{16 \pi G} - \frac{\Lambda }{8 \pi  G}+\frac{\gamma }{4 \pi  G r} \,, \label{rho_P_ra_1}\\
    \rho - |P_t| &=& F'(r) \left(\frac{b'(r)}{16 \pi G r}+\frac{3 b(r)}{16 \pi  G r^2}-\frac{1}{4 \pi  G r}\right) + \left(\frac{b(r)}{8 \pi  G r} - \frac{1}{8 \pi G}\right) F''(r) \nonumber \\ 
    && +\frac{f(r)}{16 \pi G}-\frac{\Lambda }{8 \pi  G}+\frac{\gamma }{4 \pi  G r} - \bigg| F'(r) \left(-\frac{b'(r)}{16 \pi  G r}-\frac{3 b(r)}{16 \pi  G r^2}+\frac{1}{4 \pi  G r}\right) \nonumber \\
    && + F(r) \left(\frac{b'(r)}{16 \pi  G r^2}+\frac{b(r)}{16 \pi  G r^3} \right) + \left(\frac{1}{8 \pi G} - \frac{b(r)}{8 \pi  G r}\right) F''(r) \nonumber \\
    && -\frac{f(r)}{16 \pi  G}+\frac{\Lambda }{8 \pi G}-\frac{\gamma }{8 \pi  G r} \bigg| \,, \label{rho_P_ta_1}
\end{eqnarray}

\begin{eqnarray}
    \rho + 2P_t &=& F'(r) \left(-\frac{b'(r)}{16 \pi  G r}-\frac{3 b(r)}{16 \pi  G r^2}+\frac{1}{4 \pi  G r}\right)+F(r) \left(\frac{b'(r)}{8 \pi  G r^2} + \frac{b(r)}{8 \pi  G r^3}\right) \nonumber \\
    && + \left(\frac{1}{8 \pi G} - \frac{b(r)}{8 \pi  G r}\right) F''(r) - \frac{f(r)}{16 \pi  G} + \frac{\Lambda }{8 \pi  G} \,, \label{rho_2P_t_1}\\
    \rho + P_r + 2P_t &=& F'(r) \left(-\frac{b'(r)}{16 \pi  G r}-\frac{7 b(r)}{16 \pi  G r^2}+\frac{1}{2 \pi  G r}\right)+\frac{F(r) b'(r)}{4 \pi  G r^2} \nonumber \\ 
    && +\left(\frac{1}{8 \pi  G}-\frac{b(r)}{8 \pi  G r}\right) F''(r)-\frac{f(r)}{8 \pi  G}+\frac{\Lambda }{4 \pi  G}-\frac{\gamma }{4 \pi  G r} \,. \label{rho_P_t_2P_t_1}
\end{eqnarray}
In this case, the results are plotted in Fig.\ref{E1} and summarized in Table\,\ref{mod1}. From Fig.\ref{E1}, we choose $\alpha_{1}=3.0>0$ as an example. The energy density is positive for $r \in (0, 0.40] \cup [1.10,\infty)$. The first NEC term $\rho + P_r$ is positive for $r\in (0,0.39) \cup (1.05,\infty)$, while the second NEC term $\rho + P_t$ is positive for $r\in(0.00,0.20) \cup (0.40,\infty)$. This shows that NEC and hence WEC are satisfied for $r\in [1.10,\,\infty)$. The first DEC term
$\rho - |P_r|>0$ for all $r\in[0.83,0.39] \cup [1.17, \infty)$, while the second DEC term $\rho - |P_t|>0$ for all $r\in[0.0,0.20] \cup [0.39,\,0.40]\cup [1.09, \infty)$. This shows the DEC is satisfied for $r\in [1.17,\,\infty)$. However, we find that the SEC is partially violated since the third SEC term $\rho+P_{r}+2P_{t}<0$ for $r\in [1.18,\,\infty)$.

\subsection{$\Phi(r) = \frac{\gamma_{1}}{r}$}
With the shape function from Eq.~(\ref{shap1}) and $\Phi(r) = \frac{\gamma_{1}}{r}$, we find
\begin{eqnarray}
    \rho &=& F'(r) \left(\frac{b'(r)}{16 \pi  G r}-\frac{\gamma _1 b(r)}{8 \pi  G r^3}+\frac{3 b(r)}{16 \pi G r^2}+\frac{\gamma _1}{8 \pi  G r^2}-\frac{1}{4 \pi  G r}\right) \nonumber \\
    && +F(r) \left(\frac{\gamma _1 b'(r)}{16 \pi  G r^3}-\frac{\gamma _1^2 b(r)}{8 \pi  G r^5}-\frac{\gamma _1 b(r)}{16 \pi G r^4} + \frac{\gamma _1^2}{8 \pi  G r^4} \right) + \left(\frac{b(r)}{8 \pi  G r}-\frac{1}{8 \pi G} \right) F''(r) \nonumber \\
    && +\frac{f(r)}{16 \pi  G}-\frac{\Lambda }{8 \pi  G}+\frac{\gamma }{4 \pi  G r} \,, \label{rho_2}
\end{eqnarray}
\begin{eqnarray}
    P_r &=& F(r) \left(-\frac{\gamma _1 b'(r)}{16 \pi  G r^3}+\frac{b'(r)}{8 \pi  G r^2}+\frac{\gamma _1^2 b(r)}{8 \pi  G r^5} + \frac{5 \gamma _1 b(r)}{16 \pi  G r^4} - \frac{b(r)}{8 \pi  G r^3 } - \frac{\gamma_1^2 }{8 \pi  G r^4} - \frac{\gamma _1}{4 \pi  G r^3}\right) \nonumber \\
    && +F'(r) \left(\frac{\gamma _1 b(r)}{8 \pi G r^3} - \frac{b(r)}{4 \pi  G r^2}-\frac{\gamma _1}{8 \pi  G r^2} + \frac{1}{4 \pi G r}\right) - \frac{f(r)}{16 \pi  G}+\frac{\Lambda }{8 \pi  G}-\frac{\gamma }{4 \pi  G r} \,, \label{P_r_2}\\
    P_t &=& F'(r) \left(-\frac{b'(r)}{16 \pi  G r}+\frac{\gamma _1 b(r)}{8 \pi  G r^3}-\frac{3 b(r)}{16 \pi G r^2} - \frac{\gamma _1}{8 \pi  G r^2}+\frac{1}{4 \pi  G r}\right) \nonumber \\
    && +F(r) \left(\frac{b'(r)}{16 \pi G r^2}-\frac{\gamma _1 b(r)}{8 \pi  G r^4}+\frac{b(r)}{16 \pi  G r^3} + \frac{\gamma _1}{8 \pi G r^3}\right)+\left(\frac{1}{8 \pi  G}-\frac{b(r)}{8 \pi  G r}\right) F''(r) \nonumber \\
    &&- \frac{f(r)}{16 \pi G} + \frac{\Lambda }{8 \pi  G}-\frac{\gamma }{8 \pi  G r} \,. \label{P_t_2}
\end{eqnarray}
The combinations of Eqs.~(\ref{rho_2} - \ref{P_t_2}) yield the following relations among $\rho, P_r, \text{ and } P_t$:
\begin{eqnarray}
    \rho + P_r &=& F'(r) \left(\frac{b'(r)}{16 \pi  G r}-\frac{b(r)}{16 \pi  G r^2}\right)+F(r) \left(\frac{b'(r)}{8 \pi G r^2} + \frac{\gamma _1 b(r)}{4 \pi  G r^4} - \frac{b(r)}{8 \pi  G r^3}-\frac{\gamma _1}{4 \pi  G   r^3}\right)\nonumber \\
    && + \left(\frac{b(r)}{8 \pi  G r} - \frac{1}{8 \pi  G}\right) F''(r) \,, \label{rho_P_r_2},
\end{eqnarray}
\begin{eqnarray}
    \rho + P_t &=& F(r) \left(\frac{\gamma _1 b'(r)}{16 \pi  G r^3}+\frac{b'(r)}{16 \pi  G r^2}-\frac{\gamma_1^2 b(r)}{8 \pi  G r^5}-\frac{3 \gamma _1 b(r)}{16 \pi  G r^4}+\frac{b(r)}{16 \pi  G r^3}+\frac{\gamma   _1^2}{8 \pi  G r^4}+\frac{\gamma _1}{8 \pi  G r^3}\right) \nonumber \\
    && +\frac{\gamma }{8 \pi  G r} \,, \label{rho_P_t_2}
\end{eqnarray}
\begin{eqnarray}
    \rho - |P_r| &=& -\bigg| -\frac{\gamma }{4 G \pi  r}+\frac{\Lambda }{8 G \pi }-\frac{f(r)}{16 G \pi } + F(r) \bigg(\frac{b(r) \gamma _1^2}{8 G \pi  r^5}-\frac{\gamma _1^2}{8 G \pi  r^4}+\frac{5 b(r) \gamma_1}{16 G \pi  r^4} \nonumber \\
    && -\frac{b'(r) \gamma _1}{16 G \pi  r^3} - \frac{\gamma _1}{4 G \pi r^3}-\frac{b(r)}{8 G \pi  r^3} + \frac{b'(r)}{8 G \pi  r^2}\bigg) + \bigg(\frac{\gamma _1 b(r)}{8 G  \pi  r^3}-\frac{b(r)}{4 G \pi  r^2} -\frac{\gamma _1}{8 G \pi  r^2}\nonumber \\
    && +\frac{1}{4 G \pi  r}\bigg) F'(r)\bigg| +F'(r) \left(\frac{b'(r)}{16 \pi  G r}-\frac{\gamma _1 b(r)}{8 \pi  G r^3}+\frac{3 b(r)}{16 \pi  G r^2}+\frac{\gamma _1}{8 \pi  G r^2}-\frac{1}{4 \pi  G r}\right)\nonumber \\
    && +F(r) \left(\frac{\gamma _1 b'(r)}{16 \pi  G r^3}-\frac{\gamma _1^2 b(r)}{8 \pi  G r^5}-\frac{\gamma _1  b(r)}{16 \pi  G r^4}+\frac{\gamma _1^2}{8 \pi  G r^4}\right)+\left(\frac{b(r)}{8 \pi G r} - \frac{1}{8 \pi  G}\right) F''(r) \nonumber \\
    &&+ \frac{f(r)}{16 \pi  G}-\frac{\Lambda }{8 \pi  G}+\frac{\gamma}{4 \pi  G r} \,, \label{rho_P_ra_2}\\
    \rho - |P_t| &=& -\bigg| -\frac{\gamma }{8 G \pi  r}+\frac{\Lambda }{8 G \pi }-\frac{f(r)}{16 G \pi }+F(r)   \left(-\frac{\gamma _1 b(r)}{8 G \pi  r^4} + \frac{b(r)}{16 G \pi  r^3}+\frac{\gamma _1}{8 G \pi r^3} + \frac{b'(r)}{16 G \pi  r^2}\right) \nonumber \\
    && +\left(\frac{\gamma _1 b(r)}{8 G \pi  r^3}-\frac{3 b(r)}{16 G \pi r^2} - \frac{\gamma _1}{8 G \pi  r^2}-\frac{b'(r)}{16 G \pi  r}+\frac{1}{4 G \pi r}\right) F'(r)+\bigg(\frac{1}{8 G \pi } \nonumber \\
    && -\frac{b(r)}{8 G \pi  r}\bigg) F''(r)\bigg| +F'(r) \bigg(\frac{b'(r)}{16 \pi  G r}-\frac{\gamma _1 b(r)}{8 \pi G r^3}+\frac{3 b(r)}{16 \pi G r^2}+\frac{\gamma _1}{8 \pi  G r^2}-\frac{1}{4 \pi  G r}\bigg) \nonumber \\
    && + F(r) \bigg( \frac{ \gamma_1 b'(r) }{16 \pi  G r^3} - \frac{\gamma _1^2 b(r)}{8 \pi  G r^5}-\frac{\gamma _1 b(r)}{16 \pi G r^4}+\frac{\gamma _1^2}{8 \pi  G r^4}\bigg) + \bigg(\frac{b(r)}{8 \pi  G r}-\frac{1}{8 \pi   G}\bigg) F''(r) \nonumber \\
    && +\frac{f(r)}{16 \pi  G}-\frac{\Lambda }{8 \pi  G} + \frac{\gamma }{4 \pi  G r}\,, \label{rho_P_ta_2}\\
    \rho + 2P_t &=& F'(r) \left(-\frac{b'(r)}{16 \pi  G r}+\frac{\gamma _1 b(r)}{8 \pi  G r^3}-\frac{3 b(r)}{16 \pi  G r^2}-\frac{\gamma _1}{8 \pi  G r^2}+\frac{1}{4 \pi  G r}\right) \nonumber \\
    && +F(r) \left(\frac{\gamma _1  b'(r)}{16 \pi  G r^3}+\frac{b'(r)}{8 \pi  G r^2}-\frac{\gamma _1^2 b(r)}{8 \pi  G r^5}-\frac{5   \gamma _1 b(r)}{16 \pi  G r^4}+\frac{b(r)}{8 \pi  G r^3}+\frac{\gamma _1^2}{8 \pi  G   r^4}+\frac{\gamma _1}{4 \pi  G r^3}\right) \nonumber \\
    && +\left(\frac{1}{8 \pi  G}-\frac{b(r)}{8 \pi  G   r}\right) F''(r)-\frac{f(r)}{16 \pi  G}+\frac{\Lambda }{8 \pi  G} \,, \label{rho_2P_t_2}\\
    \rho + P_r + 2P_t &=& F'(r) \left(-\frac{b'(r)}{16 \pi  G r}+\frac{\gamma _1 b(r)}{4 \pi  G r^3}-\frac{7 b(r)}{16 \pi  G r^2}-\frac{\gamma _1}{4 \pi  G r^2}+\frac{1}{2 \pi  G r}\right)+\frac{F(r) b'(r)}{4 \pi  G  r^2} \nonumber \\
    && +\left(\frac{1}{8 \pi  G}-\frac{b(r)}{8 \pi  G r}\right) F''(r)-\frac{f(r)}{8 \pi G}+\frac{\Lambda }{4 \pi  G}-\frac{\gamma }{4 \pi  G r} \,.\label{rho_P_t_2P_t_2}
\end{eqnarray}
\begin{figure}
    \centering
    \includegraphics[width = 8 cm]{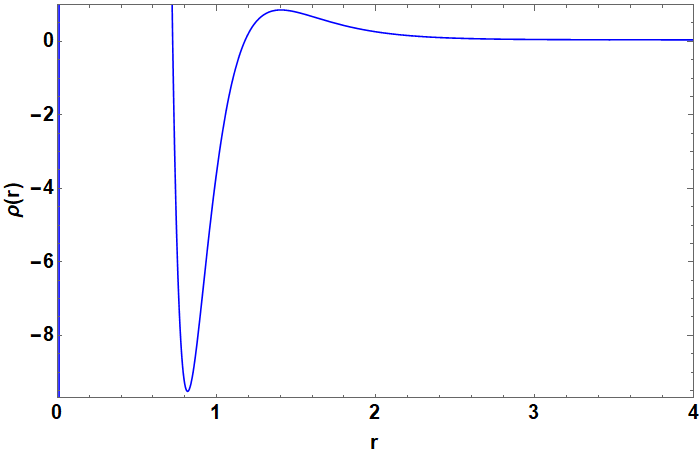}
    \includegraphics[width = 8 cm]{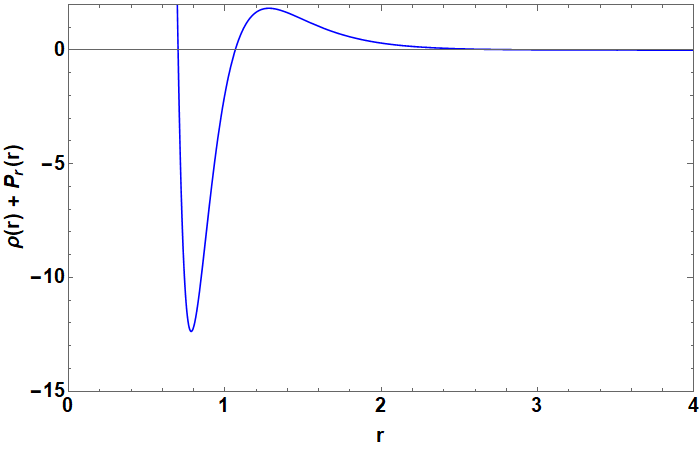}
    \includegraphics[width = 8 cm]{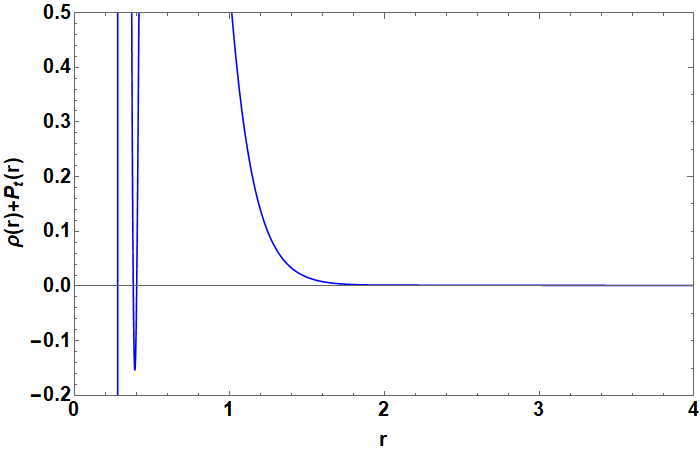}
    \includegraphics[width = 8 cm]{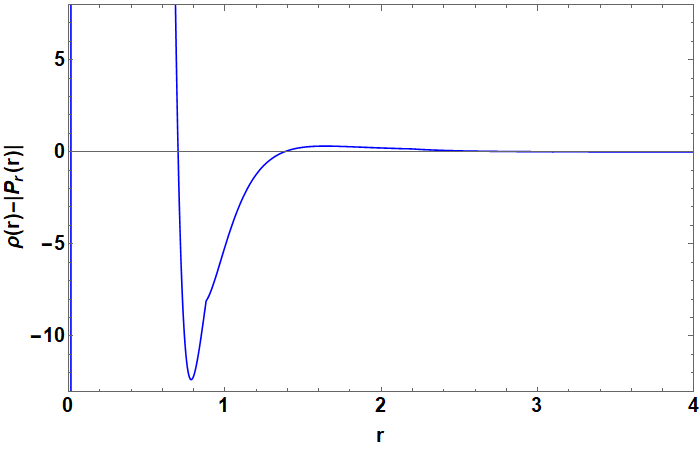}
    \includegraphics[width = 8 cm]{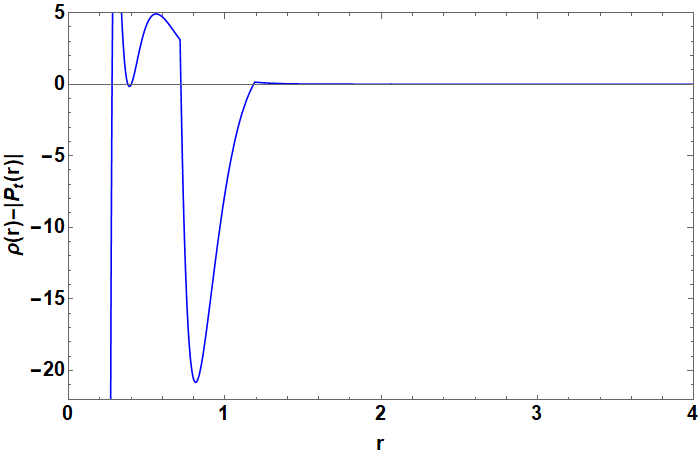}
    \includegraphics[width = 8 cm]{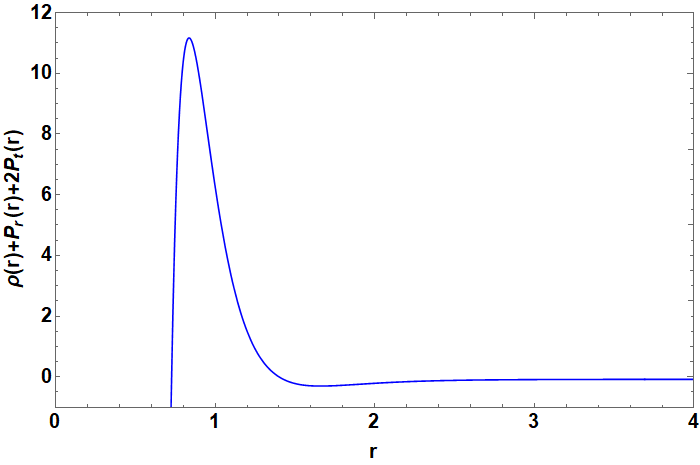}
    \caption{Figures illustrate the variation of $\rho$, $\rho + P_r$, $\rho + P_t$, $\rho - |P_r|$, $\rho - |P_t|$, $\rho + P_r + 2 P_t$ as a function of $r$ with $\Phi(r) = \gamma_1 / r$. Here we have used $\alpha_{1} = 3, n = 2, \alpha = 3.0, r_0 = 1 \text{ and } \gamma_1 = 1.0$.} 
    \label{E2}
\end{figure}
\begin{table}[!h]
	\centering
	\caption{Table shows a summary of energy/pressure conditions for $\Phi(r) = \gamma_1 / r$,\,$n=2$, $\alpha = 3.0$, $ r_0 = 1, \gamma_1 = 1.0, \gamma = 0.1, \text{ and } \Lambda = -1.0$.}
	\begin{tabular}{|c|c|l|l|}
		\hline\hline
		No.& Terms& $\alpha_{1} < 0$& $\alpha_{1} > 0$ \\
		\hline\hline
		1 & $\rho$ & $\geq 0$, for $r\in (0,0.01] \cup [1.03,1.42] $& $\geq 0$, for $r\in[0.01, 1.03] \cup [1.40, \infty)$\\
		&       & \quad $\cup [2.41,\infty)$     & \\
		&       & $<0$, for $r \in (0.01, 1.03) \cup (1.42,2.41)$ &  $< 0 $, for $r \in (0,0.01) \cup (1.03,1.40)$ \\
		\cline{1-4}
		2 & $\rho+P_{r}$ & $\geq0$, for $r\in[1.02, 1.17]$ & $\geq 0$, for $r \in (0,1.02) \cup (1.17, \infty)$ \\
		&       & $<0$, for $r \in (0.00,1.02) \cup (1.17, \infty)$ &  $<0$, for $r\in[1.02, 1.17]$ \\
		\cline{1-4}
		3 & $\rho+P_{t}$ & $\geq0$, for $r\in(0.00,0.93] \cup [1.33, \infty)$& $\geq0$, for $r\in(0.93,\infty)$\\
		&       & $<0$, for $r\in(0.93,1.33)$ &  $<0$, for $r< 0.93$ \\
		\cline{1-4}
		4 & $\rho+P_{r}+2P_{t}$ & $\geq 0$, for $r \in [1.00, 1.50]$ & $\geq 0$, for $r \in (0.00, 1.00] \cup [1.71,2.00]$\\
		&       & $<0$, for $r \in (0.00,1.00) \cup (1.50, \infty)$   & $<0$, for $r \in (1.00,1.71) \cup (2.00, \infty) $\\
		\cline{1-4}
		5 & $\rho-|P_{r}|$ & $\geq0$, for $r\in[1.07,1.31]$ & $\geq 0$, for $r\in[0.12, 1.06] \cup [1.49, \infty)$\\
		&       & $<0$, for $r \in (0.00, 1.07) \cup (1.31, \infty)$ &  $<0$, for $r \in (0.00, 0.12) \cup (1.06, 1.49)$\\
		\cline{1-4}
		6 & $\rho-|P_{t}|$ & $\geq0$, for $r\in [1.22, 1.40] \cup [2.49, \infty)$& $\geq0$, for $r\in [0.91, 1.00] \cup [1.39, \infty)$\\
		&       & $<0$, for $r \in (0.00, 1.22) \cup (1.40,2.49)$ &  $<0$, for $r \in (0.00, 0.91) \cup (1.22, 1.40)$ \\
		\hline
	\end{tabular}\label{mod2}
\end{table}
In the second case, the results are plotted in Fig.\ref{E2} and summarized in Table\,\ref{mod2}. From Fig.\ref{E2}, we choose $\alpha_{1}=3.0>0$ as an example. The energy density is positive for $r\in[0.01, 1.03] \cup [1.40, \infty)$. The first NEC term $\rho + P_r$ is positive for $r \in (0.00,1.02) \cup (1.17, \infty)$, while the second NEC term $\rho + P_t$ is positive for $r\in(0.93,\infty)$. This shows that NEC and hence WEC are satisfied for $r\in [1.40,\,\infty)$. The first DEC term
$\rho - |P_r|>0$ for all $r\in[0.12, 1.06] \cup [1.49, \infty)$, while the second DEC term $\rho - |P_t|>0$ for all $r\in [0.91, 1.00] \cup [1.39, \infty)$. This shows the DEC is satisfied for $r\in [1.49,\,\infty)$. However, we find that the SEC is partially violated since the third SEC term $\rho+P_{r}+2P_{t}<0$ for $r \in (1.00,1.71) \cup (2.00, \infty) $.

\subsection{$\Phi(r) = \log \left( 1 + \frac{\gamma_{2}}{r} \right)$}
With the shape function from Eq.~(\ref{shap1}) and $\Phi(r) = \log \left( 1 + \frac{\gamma_{2}}{r} \right)$, we find
\begin{eqnarray}
    \rho &=& F'(r) \left(\frac{b'(r)}{16 \pi  G r}-\frac{\gamma _2 b(r)}{8 \pi  G r^3 \left(\frac{\gamma_2}{r} + 1\right)}+\frac{3 b(r)}{16 \pi  G r^2}+\frac{\gamma _2}{8 \pi  G r^2 \left(\frac{\gamma_2}{r} + 1\right)}-\frac{1}{4 \pi  G r}\right) \nonumber \\ 
    && +F(r) \left(\frac{\gamma _2 b'(r)}{16 \pi  G r^3 \left(\frac{\gamma_2}{r} + 1\right)}-\frac{\gamma _2 b(r)}{16 \pi  G r^4 \left(\frac{\gamma_2}{r} + 1\right)}\right) + \left(\frac{b(r)}{8 \pi  G r}-\frac{1}{8 \pi  G}\right) F''(r) \nonumber \\
    && + \frac{f(r)}{16 \pi  G} - \frac{\Lambda }{8 \pi  G} + \frac{\gamma }{4 \pi  G r} \,, \label{rho_3}
\end{eqnarray}
\begin{figure}[!h]
    \centering
    \includegraphics[width = 8 cm]{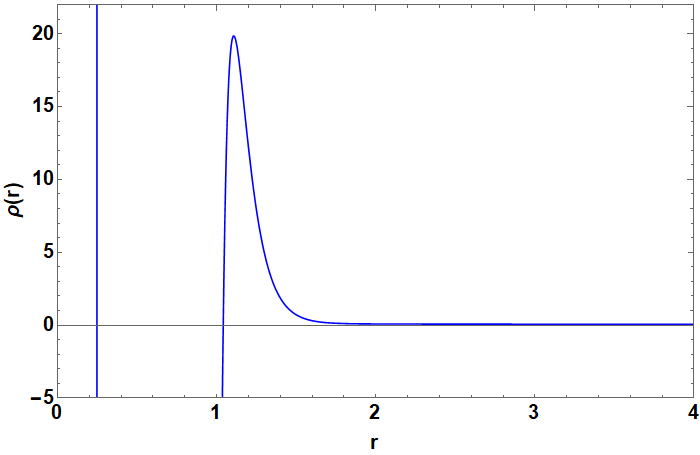}
    \includegraphics[width = 8 cm]{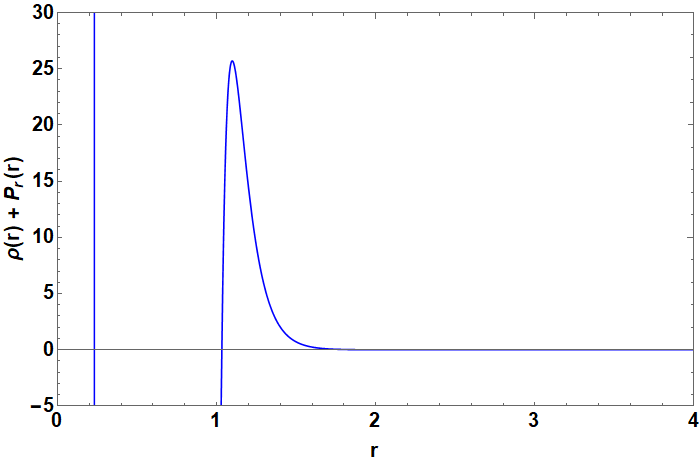}
    \includegraphics[width = 8 cm]{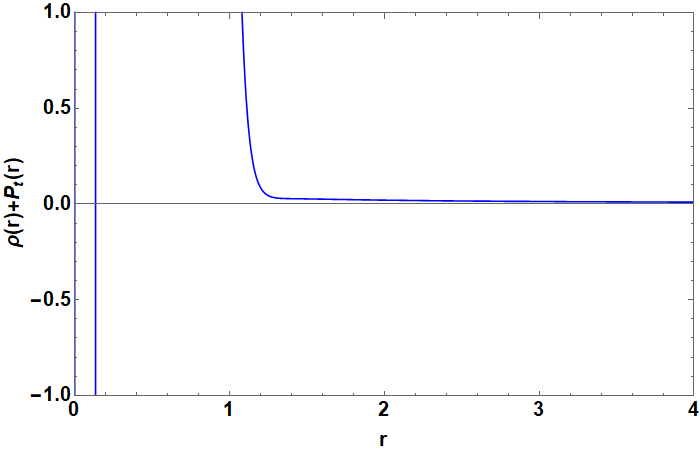}
    \includegraphics[width = 8 cm]{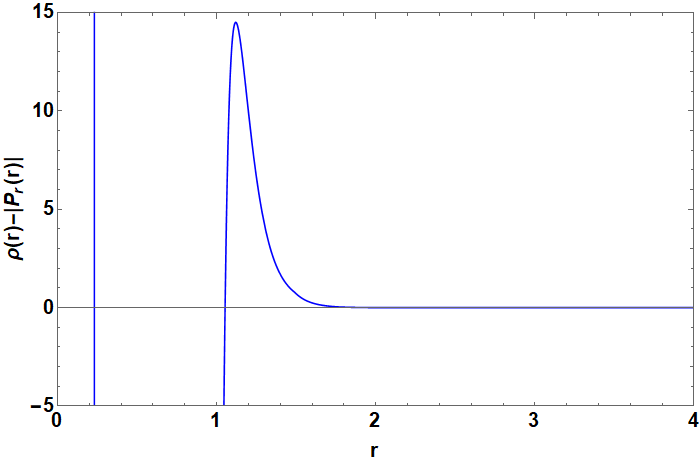}
    \includegraphics[width = 8 cm]{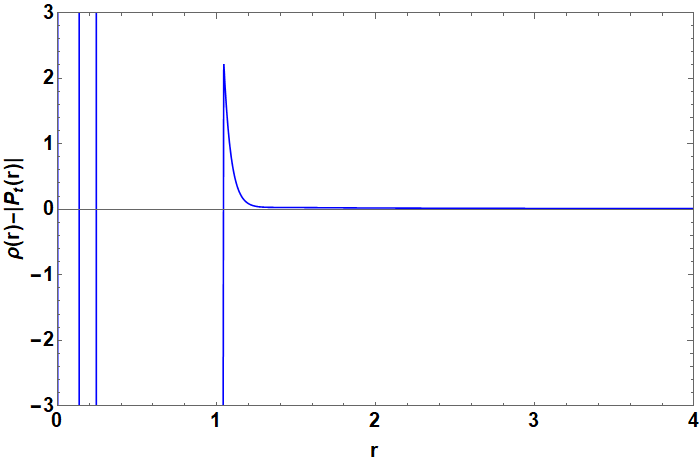}
    \includegraphics[width = 8 cm]{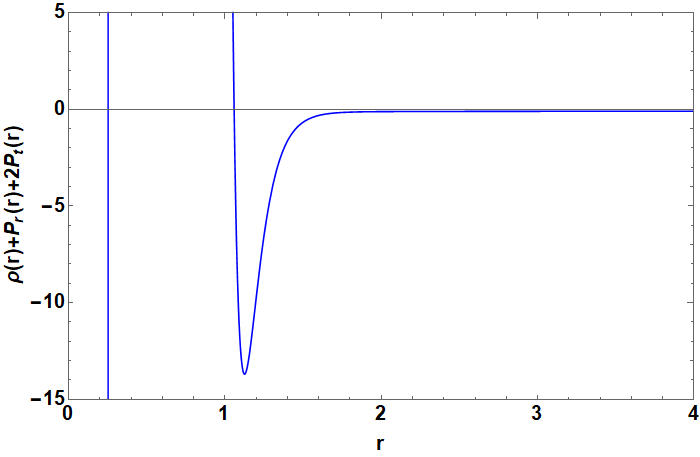}
    \caption{Figures illustrate the variation of $\rho$, $\rho + P_r$, $\rho + P_t$, $\rho - |P_r|$, $\rho - |P_t|$, $\rho + P_r + 2 P_t$ as a function of $r$ with $\Phi(r) = \log (1 + \frac{\gamma_2}{r})$. Here we have used $\alpha_{1} = 1, n = 2, \alpha = 10.0, r_0 = 1 \text{ and } \gamma_2 = 0.15$.} 
    \label{E3}
\end{figure}
\begin{table}[!h]
	\centering
	\caption{Table shows a summary of energy/pressure conditions for $\Phi(r) = \log (1 + \gamma_2 / r)$,\,$n=2$, $\alpha = 10.0$, $ r_0 = 1, \gamma_2 = 0.15, \gamma = 1.0, \text{ and } \Lambda = -1.0$.}
	\begin{tabular}{|c|c|l|l|}
		\hline\hline
		No.& Terms& $\alpha_{1} < 0$& $\alpha_{1} > 0$ \\
		\hline\hline
		1 & $\rho$ & $\geq 0$, for $r\in [0.22,1.02] \cup [1.67, \infty) $& $\geq 0$, for $r\in(0.00, 0.24] \cup [1.04, \infty)$\\
		&       & $<0$, for $r \in (0.00, 0.22) \cup (1.02, 1.67)$ &  $< 0 $, for $r \in (0.24, 1.04)$ \\
		\cline{1-4}
		2 & $\rho+P_{r}$ & $\geq0$, for $r\in[0.22, 1.02]$ & $\geq 0$, for $r \in (0,0.21] \cup [1.02, 2.11]$ \\
		&       & $<0$, for $r \in (0.00,0.22) \cup (1.02, \infty)$ &  $<0$, for $r\in (0.21, 1.02) \cup (2.11, \infty)$ \\
		\cline{1-4}
		3 & $\rho+P_{t}$ & $\geq0$, for $r\in(0.00,0.14] \cup [1.19, \infty)$& $\geq0$, for $r\in(0.13, \infty)$\\
		&       & $<0$, for $r\in(0.14,1.19)$ &  $<0$, for $r< 0.23$ \\
		\cline{1-4}
		4 & $\rho+P_{r}+2P_{t}$ & $\geq 0$, for $r \in (0.00, 0.23] \cup [1.07, 1.63]$ & $\geq 0$, for $r \in [0.14, 1.04]$\\
		&       & $<0$, for $r \in (0.23, 1.07) \cup (1.63, \infty)$   & $<0$, for $r \in (0.00, 0.14) \cup (1.04, \infty)$\\
		\cline{1-4}
		5 & $\rho-|P_{r}|$ & $\geq0$, for $r\in[0.27, 1.02] $ & $\geq 0$, for $r\in(0.00, 0.22] \cup [1.03, 2.11]$\\
		&       & $<0$, for $r \in (0.00, 0.27) \cup (1.02, \infty)$ &  $<0$, for $r \in (0.22, 1.03) \cup (2.11, \infty)$\\
		\cline{1-4}
		6 & $\rho-|P_{t}|$ & $\geq0$, for $r\in [1.69, \infty)$& $\geq0$, for $r\in [0.15, 0.22] \cup [1.25, \infty)$\\
		&       & $<0$, for $r \in (0.00, 1.22) \cup (1.40,2.49)$ &  $<0$, for $r <1.69$ \\
		\hline
	\end{tabular}\label{mod3}
\end{table}
\begin{eqnarray}
    P_r &=& F(r) \left(-\frac{\gamma _2 b'(r)}{16 \pi  G r^3 \left(\frac{\gamma _2}{r} + 1\right)}+\frac{b'(r)}{8\pi  G r^2} + \frac{5 \gamma _2 b(r)}{16 \pi  G r^4 \left(\frac{\gamma_2}{r} + 1\right)}-\frac{b(r)}{8 \pi  G r^3}-\frac{\gamma _2}{4 \pi  G r^3 \left(\frac{\gamma_2}{r} + 1\right)}\right) \nonumber \\
    && + F'(r) \left(\frac{\gamma _2 b(r)}{8 \pi  G r^3 \left(\frac{\gamma   _2}{r} + 1\right)}-\frac{b(r)}{4 \pi  G r^2}-\frac{\gamma _2}{8 \pi  G r^2 \left(\frac{\gamma_2}{r} + 1\right)} + \frac{1}{4 \pi  G r}\right)\nonumber \\
    &&-\frac{f(r)}{16 \pi  G}+\frac{\Lambda }{8 \pi G}-\frac{\gamma }{4 \pi  G r} \,, \label{P_r_3}
\end{eqnarray}
\begin{eqnarray}
    P_t &=& F'(r) \left(-\frac{b'(r)}{16 \pi  G r}+\frac{\gamma _2 b(r)}{8 \pi  G r^3 \left(\frac{\gamma_2}{r} + 1\right)} - \frac{3 b(r)}{16 \pi  G r^2}-\frac{\gamma _2}{8 \pi  G r^2 \left(\frac{\gamma_2}{r} + 1\right)} + \frac{1}{4 \pi  G r}\right) \nonumber \\
    && + F(r) \left(\frac{b'(r)}{16 \pi  G r^2}-\frac{\gamma_2 b(r)}{8 \pi  G r^4 \left(\frac{\gamma _2}{r} + 1\right)}+\frac{b(r)}{16 \pi  G r^3}+\frac{\gamma_2}{8 \pi  G r^3 \left(\frac{\gamma_2}{r} + 1\right)}\right)\nonumber \\
    && + \left(\frac{1}{8 \pi  G}-\frac{b(r)}{8 \pi  G r}\right) F''(r)  - \frac{f(r)}{16 \pi G} + \frac{\Lambda }{8 \pi  G}-\frac{\gamma }{8 \pi G r} \,. \label{P_t_3}
\end{eqnarray}
The combinations of Eqs.~(\ref{rho_3} - \ref{P_t_3}) yield the following relations among $\rho, P_r, \text{ and } P_t$:
\begin{eqnarray}
    \rho + P_r &=& F'(r) \left(\frac{b'(r)}{16 \pi  G r}-\frac{b(r)}{16 \pi  G r^2}\right) + F(r) \bigg(\frac{b'(r)}{8   \pi  G r^2}+\frac{\gamma _2 b(r)}{4 \pi  G r^4 \left(\frac{\gamma _2}{r} + 1\right)} - \frac{b(r)}{8\pi G r^3} \nonumber \\
    && -\frac{\gamma _2}{4 \pi  G r^3 (\frac{\gamma  _2}{r}+1)}\bigg) + \left(\frac{b(r)}{8 \pi  G r}-\frac{1}{8 \pi  G}\right) F''(r) \,, \label{rho_P_r_3}\\
    \rho + P_t &=& F(r) \bigg(\frac{\gamma _2 b'(r)}{16 \pi  G r^3 \left(\frac{\gamma _2}{r}+1\right)}+\frac{b'(r)}{16   \pi  G r^2}-\frac{3 \gamma _2 b(r)}{16 \pi  G r^4 \left(\frac{\gamma_2}{r}+1\right)} + \frac{b(r)}{16 \pi  G r^3} \nonumber \\
    && +\frac{\gamma _2}{8 \pi  G r^3 \left(\frac{\gamma   _2}{r}+1\right)}\bigg)+\frac{\gamma }{8 \pi  G r} \,, \label{rho_P_t_3}
\end{eqnarray}
\begin{eqnarray}
    \rho - |P_r| &=& -\bigg| -\frac{\gamma }{4 G \pi  r}+\frac{\Lambda }{8 G \pi }-\frac{f(r)}{16 G \pi } + F(r)   \bigg(\frac{5 \gamma _2 b(r)}{16 G \pi  r^4 \left(\frac{\gamma _2}{r}+1\right)} - \frac{b(r)}{8 G   \pi  r^3} \nonumber \\
    && -\frac{\gamma _2 b'(r)}{16 G \pi  r^3 \left(\frac{\gamma _2}{r}+1\right)}+\frac{b'(r)}{8 G \pi  r^2}-\frac{\gamma _2}{4 G \pi  r^3 \left(\frac{\gamma   _2}{r}+1\right)}\bigg) + \bigg(\frac{\gamma _2 b(r)}{8 G \pi  r^3 \left(\frac{\gamma _2}{r}+1\right)} \nonumber \\ 
    && -\frac{b(r)}{4 G \pi  r^2}+\frac{1}{4 G \pi  r}-\frac{\gamma _2}{8 G \pi  r^2 \left(\frac{\gamma _2}{r}+1\right)}\bigg) F'(r)\bigg| +F'(r) \bigg(\frac{b'(r)}{16 \pi  G   r} \nonumber \\
    && -\frac{\gamma _2 b(r)}{8 \pi  G r^3 \left(\frac{\gamma _2}{r}+1\right)}+\frac{3 b(r)}{16 \pi  G  r^2}+\frac{\gamma _2}{8 \pi  G r^2 \left(\frac{\gamma _2}{r} + 1\right)} - \frac{1}{4 \pi  G r}\bigg)\nonumber \\
    && +F(r) \left(\frac{\gamma _2 b'(r)}{16 \pi  G r^3 \left(\frac{\gamma   _2}{r} + 1\right)} - \frac{\gamma _2 b(r)}{16 \pi  G r^4 \left(\frac{\gamma   _2}{r} + 1\right)}\right) + \left(\frac{b(r)}{8 \pi  G r}-\frac{1}{8 \pi  G}\right)   F''(r) \nonumber \\
    && + \frac{f(r)}{16 \pi  G}-\frac{\Lambda }{8 \pi  G}+\frac{\gamma }{4 \pi  G r} \,,\label{rho_P_ra_3}
\end{eqnarray}
\begin{eqnarray}
    \rho - |P_t| &=& -\bigg| -\frac{\gamma }{8 G \pi r}+\frac{\Lambda }{8 G \pi }-\frac{f(r)}{16 G \pi }+F(r)   \bigg(-\frac{\gamma _2 b(r)}{8 G \pi  r^4 \left(\frac{\gamma _2}{r} + 1\right)}+\frac{b(r)}{16 G   \pi  r^3}+\frac{b'(r)}{16 G \pi  r^2} \nonumber \\
    &&+\frac{\gamma _2}{8 G \pi  r^3 \left(\frac{\gamma  _2}{r} + 1\right)}\bigg) + \bigg(\frac{\gamma _2 b(r)}{8 G \pi  r^3 \left(\frac{\gamma _2}{r}+1\right)}-\frac{3 b(r)}{16 G \pi  r^2} - \frac{b'(r)}{16 G \pi  r}+\frac{1}{4 G \pi   r} \nonumber \\ 
    &&- \frac{\gamma _2}{8 G \pi  r^2 \left(\frac{\gamma_2}{r} + 1\right)}\bigg) F'(r) + \left(\frac{1}{8 G \pi }-\frac{b(r)}{8 G \pi  r}\right) F''(r)\bigg| +F'(r)   \bigg(\frac{b'(r)}{16 \pi  G r} \nonumber \\
    && - \frac{\gamma _2 b(r)}{8 \pi  G r^3 \left(\frac{\gamma_2}{r}+1\right)}+\frac{3 b(r)}{16 \pi  G r^2}+\frac{\gamma _2}{8 \pi  G r^2 \left(\frac{\gamma_2}{r}+1\right)}-\frac{1}{4 \pi  G r}\bigg)\nonumber \\
    && + F(r) \left(\frac{\gamma _2 b'(r)}{16 \pi  G r^3   \left(\frac{\gamma _2}{r}+1\right)}-\frac{\gamma _2 b(r)}{16 \pi  G r^4 \left(\frac{\gamma_2}{r} + 1\right)}\right)+\left(\frac{b(r)}{8 \pi  G r}-\frac{1}{8 \pi  G}\right) F''(r) \nonumber \\ 
    && +\frac{f(r)}{16 \pi  G}-\frac{\Lambda }{8 \pi  G}+\frac{\gamma }{4 \pi  G r} \,, \label{rho_P_t_3}
\end{eqnarray}
\begin{eqnarray}
    \rho + 2P_t &=& F'(r) \left(-\frac{b'(r)}{16 \pi  G r}+\frac{\gamma _2 b(r)}{8 \pi  G r^3 \left(\frac{\gamma   _2}{r}+1\right)}-\frac{3 b(r)}{16 \pi  G r^2}-\frac{\gamma _2}{8 \pi  G r^2 \left(\frac{\gamma   _2}{r} + 1\right)} + \frac{1}{4 \pi  G r}\right) \nonumber \\
    && +F(r) \left(\frac{\gamma _2 b'(r)}{16 \pi  G r^3\left(\frac{\gamma _2}{r} + 1\right)} + \frac{b'(r)}{8 \pi  G r^2}-\frac{5 \gamma _2 b(r)}{16 \pi  G r^4 \left(\frac{\gamma _2}{r}+1\right)}+\frac{b(r)}{8 \pi  G r^3}+\frac{\gamma _2}{4 \pi  G r^3 \left(\frac{\gamma _2}{r}+1\right)}\right) \nonumber \\
    &&+\left(\frac{1}{8 \pi  G}-\frac{b(r)}{8 \pi G r}\right) F''(r)-\frac{f(r)}{16 \pi  G}+\frac{\Lambda }{8 \pi  G} \,,\label{rho_2P_t_3} \\
    \rho + P_r + 2P_t &=& F'(r) \left(-\frac{b'(r)}{16 \pi  G r}+\frac{\gamma _2 b(r)}{4 \pi  G r^3 \left(\frac{\gamma   _2}{r}+1\right)}-\frac{7 b(r)}{16 \pi  G r^2}-\frac{\gamma _2}{4 \pi  G r^2 \left(\frac{\gamma   _2}{r} + 1\right)} + \frac{1}{2 \pi  G r}\right) \nonumber \\ 
    && +\frac{F(r) b'(r)}{4 \pi  G r^2}+\left(\frac{1}{8 \pi G} - \frac{b(r)}{8 \pi  G r}\right) F''(r)-\frac{f(r)}{8 \pi  G}+\frac{\Lambda }{4 \pi G} - \frac{\gamma }{4 \pi  G r} \,. \label{rho_P_t_2P_t_3}
\end{eqnarray}
In the last case, the results are plotted in Fig.\ref{E3} and summarized in Table\,\ref{mod3}. From Fig.\ref{E3}, we choose $\alpha_{1}=3.0>0$ as an example. The energy density is positive for $r\in(0.00, 0.24] \cup [1.04, \infty)$. The first NEC term $\rho + P_r$ is positive for $r \in (0,0.21] \cup [1.02, 2.11]$ and negative for $r\in (0.21, 1.02) \cup (2.11, \infty)$, while the second NEC term $\rho + P_t$ is positive for $r\in(0.13, \infty)$. This shows that NEC and hence WEC are partially violated for this model. The first DEC term
$\rho - |P_r|>0$ for $r\in(0.00, 0.22] \cup [1.03, 2.11]$ and $\rho - |P_r|<0$ for $r \in (0.22, 1.03) \cup (2.11, \infty)$, while the second DEC term $\rho - |P_t|>0$ for $r\in [0.15, 0.22] \cup [1.25, \infty)$. This shows the DEC is satisfied for $r\in [1.49,\,\infty)$ and violated for $r \in (0.22, 1.03) \cup (2.11, \infty)$. However, we find that the SEC is violated for this model.

\section{Amount of exotic matter}
\label{sec5}
In this section, we discuss the “volume integral,” which basically provides information
about the “total amount” of averaged null energy condition (ANEC) violating matter in the spacetime. This quantity is related only to $\rho$ and $P_r$, not to the transverse components. It is defined in terms of the following definite integral as \cite{Jusufi:2020rpw}
\begin{eqnarray}
\mathcal{I}_V=2\int_{r_{0}}^{\infty} \left(  \rho+P_r  \right)dV=8 \pi \int_{r_{0}}^{\infty} \left(\rho+P_r  \right)r^2  dr.\label{4}
\end{eqnarray}
\begin{figure}[!h]
    \centering
    \includegraphics[width = 10cm]{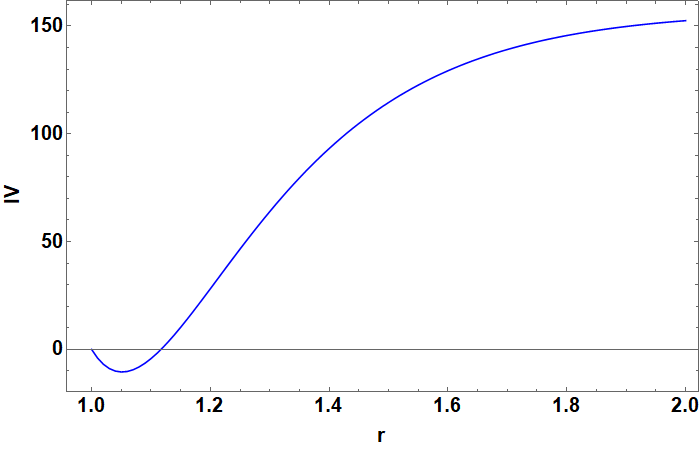}
    \caption{Figure displays the variation of $\mathcal{I}_V$ against $r=r_{\star}$ for the case of $\Phi(r) = \text{ constant}=1$ where we have set $r_{0}=1$. Here we have used $\alpha_{1} = 3.0, \alpha = 5.0, r_0 = 1.0, \gamma = 0.1, \Lambda = -1.0, \text{ and } n = 2.0$.}
    \label{IV1}
\end{figure}
Here we are going to evaluate this integral for our shape function $b(r)$. Having introduced a cut-off $r_{\star}$ such that the wormhole extends from $r_0$ to $r_{\star}$ with $r_{\star} \geq r_0$, we have instead $\mathcal{I}_V=8 \pi \int_{r_{0}}^{r_{\star}} \left(  \rho+P_r  \right)r^2  dr$. We now consider three types of the redshift function.
\begin{figure}[!h]
    \centering
    \includegraphics[width = 10cm]{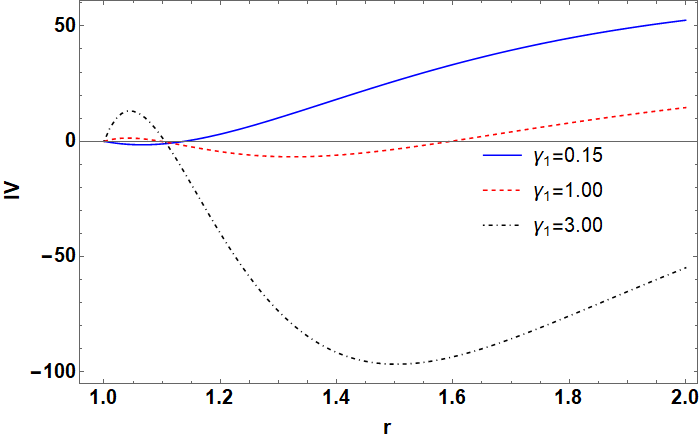}
    \caption{Figure displays the variation of $\mathcal{I}_V$ against $r=r_{\star}$ for the case of $\Phi(r) = \frac{\gamma_1}{r}$ by setting $r_{0}=1.0$. From the plots, we have used $\alpha = 3.0, \alpha_{1} = 3.0, \gamma = 0.1, \Lambda = -1.0 \text{ and } n = 2$ with various values of $\gamma_1=0.15, 1.00 \text{ and } 3.00$.}
    \label{IV2}
\end{figure}
\begin{figure}[!h]
    \centering
    \includegraphics[width = 10cm]{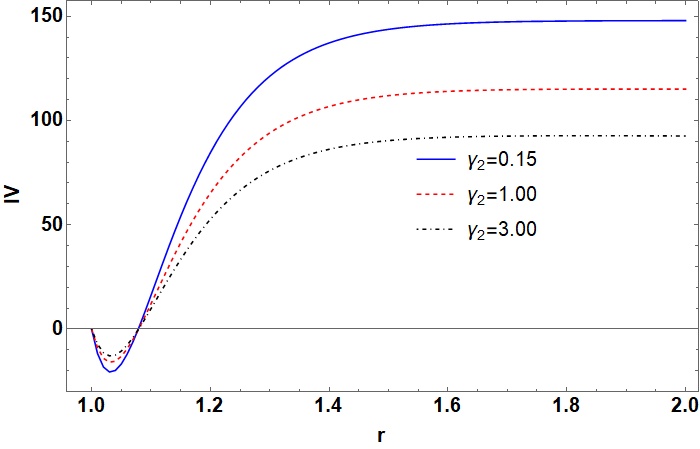}
    \caption{Figure displays the variation of $\mathcal{I}_V$ against $r=r_{\star}$ for the case of $ \Phi(r) =  \log \left( 1 + \frac{\gamma_2}{r} \right)$ with $r_{0}=1.0$. From the plots, we have used $\alpha_{1} = 1.0, \alpha = 10.0, \gamma = 1.0, \Lambda = -1.0, n = 2$. We vary three choices of $\gamma_2$ as $0.15, 1.00, \text{ and } 3.00$}
    \label{IV3}
\end{figure}
The determination of $\mathcal{I}_V$ can be straightforwardly done by substituting $\rho+P_r$ for each case into Eq.\ref{4}. Due to the limitation of space, we intentionally skip posting the full expression for ${\cal I}_{V}$ for each red-shift function model and show only the dependence of ${\cal I}_{V}$ against $r$, given by Fig.(\ref{IV1}-\ref{IV3}). 

In the first case $\Phi(r)={\rm const.}=1$, we observe from Fig.\ref{IV1} that the quantity $\mathcal{I}_V$ is negative near the wormhole throat for $r\in (1.0,\,1.13)$. In Fig.\ref{IV2}, for the second case $\Phi(r) = \frac{\gamma_1}{r}$, we find that the quantity $\mathcal{I}_V$ is negative near the wormhole throat for $r\in (1.0,\,1.10)$ using $\gamma_{1}=0.15$. However, it is negative for a larger rang of $r$ when $\gamma_{1}\gg 0.15$. Last but not the least, for the third case $ \Phi(r) =  \log \left( 1 + \frac{\gamma_2}{r} \right)$, we notice that the quantity $\mathcal{I}_V$ remains negative near the wormhole throat for $r\in (1.0,\,1.07)$ shown in Fig.\ref{IV3} where we have used three distinct values of $\gamma_2$. As a results, therefore, we demonstrate the existence of spacetime geometries containing traversable wormholes that are supported by arbitrarily
small quantities of “exotic matter”.

\section{Conclusion}

In this work, we constructed traversable wormholes in $f(R)$ gravity in the present of the dRGT massive theory. Here we have considered the function $f(R)=R+\alpha_{1} R^{n}$, where $\alpha_{1}$ and $n$ are arbitrary constants and particularly focused on $n=2$. In the present work, we have chosen the shape function of the form $b(r)=r \exp(-\alpha(r−r_{0}))$ with $\alpha$ and $r_{0}$ being an arbitrary constant and a radius of the wormhole throat, respectively. We have found that $\alpha$ affects the radius of curvature of the wormhole. Moreover, we have assume a spherically symmetric and static wormhole metric and derived field equations of the underlying description. Moreover, we have visualized the wormhole geometry using embedding diagrams. Furthermore, we have checked the null, weak, dominant and strong energy conditions at the wormhole throat with a radius $r_{0}$ invoking three types of redshift functions, $\Phi={\rm constant},\,\gamma_{1}/r,\,\log(1+\gamma_{2}/r)$ with $\gamma_{1}$ and $\gamma_{2}$ are arbitrary real constants.

In the first case $\Phi={\rm constant}=1$, the energy density is positive for $r \in (0, 0.40] \cup [1.10,\infty)$. The NEC and hence WEC are satisfied for $r\in [1.10,\,\infty)$. Our results also showed the DEC is satisfied for $r\in [1.17,\,\infty)$. However, we find that the SEC is partially violated in this model. In the second case $\Phi=\gamma_{1}/r$, the energy density is positive for $r\in[0.01, 1.03] \cup [1.40, \infty)$. We have observed that NEC and hence WEC are satisfied for $r\in [1.40,\,\infty)$. The DEC of this specific model is satisfied for $r\in [1.49,\,\infty)$. However, we have discovered that the SEC is partially violated for this model. For the last scenario $\Phi=\log(1+\gamma_{2}/r)$, we have found that the energy density is positive for $r\in(0.00, 0.24] \cup [1.04, \infty)$. Nevertheless, NEC and hence WEC, DEC are partially violated for this model. Additionally, the SEC is not satisfied in this model. However, as mentioned in Ref.\cite{Visser:1999de}, the SEC is almost abandoned.

As a final remark, using the proper choices of parameters, our results show that it is plausible to obtain traversable wormhole solutions which respect the null, weak and dominant energy conditions. In conclusion, we have checked the null, weak, dominant and strong conditions at the wormhole throat with a radius $r_{0}$, and shown that in general the classical energy conditions are violated near the wormhole throat supported by arbitrarily small quantities of yet unknown “exotic matter”. It is worth noting that this work is just one of many viable scenarios. In the context the present work, we note that one can obtain many other possible solutions for $\rho,\,P_r$ and $P_t$ by considering other choices of the functions $b(r)$, $\Phi(r)$ and $f(R)$. More specifically, other theories of $f(R)$ gravity include the Tsujikawa model \cite{Tsujikawa:2007xu}, the Hu-Sawicky model \cite{Hu:2007nk}, the Amendola-Polarski-Tsujikawa model \cite{Amendola:2006kh}, the logarithmic-corrected $R^2$ model \cite{Nojiri:2010wj,Nojiri:2003ni,Elizalde:2018now}, and the exponential model \cite{Elizalde:2010ts,Cognola:2007zu}. Moreover, the different $b(r)$ functions is also worth investigating, e.g. \cite{Samanta:2019tjb,Godani:2018blx}.

\section*{acknowledgments}
We are thankful to Prof.Mubasher Jamil for thorough review and intuitive comments on this work. TT thanks the Science Achievement Scholarship of Thailand (SAST) for financial support during his PhD study. AC is supported by the CUniverse research promotion project of Chulalongkorn University under the grant reference CUAASC. DS is supported by Thailand Research Fund (TRF) under a contract No.TRG6180014.

\end{document}